\documentclass[fleqn,usenatbib]{mnras}

\usepackage{newtxtext,newtxmath,amsmath}

\usepackage[T1]{fontenc}

\DeclareRobustCommand{\VAN}[3]{#2}
\let\VANthebibliography\thebibliography
\def\thebibliography{\DeclareRobustCommand{\VAN}[3]{##3}\VANthebibliography}

\usepackage{graphicx}	
\usepackage{amsmath}	
\usepackage[flushleft]{threeparttable}
\usepackage{multirow}
\usepackage{amsmath, bm}
\usepackage{euscript}

\usepackage{caption}
\usepackage{subcaption}
\usepackage{enumerate}

\title[Mass-ratio distribution - III]{The Frequency and Mass-Ratio Distribution of Binaries in Clusters – III: Probabilistic Generative Modelling of Six Young Open Clusters.}

\author[J. S. Alexander \& M. D. Albrow]{
Jason S. Alexander,$^{1}$\thanks{E-mail: jason.alexander@pg.canterbury.ac.nz} and 
Michael D. Albrow$^{1}$\thanks{E-mail: michael.albrow@canterbury.ac.nz}
\\
$^{1}$School of Physical and Chemical Sciences, University of Canterbury, Private Bag 4800, Christchurch, New Zealand
}

\date{Accepted XXX. Received YYY; in original form ZZZ}

\pubyear{2024}

\begin{document}
\label{firstpage}
\pagerange{\pageref{firstpage}--\pageref{lastpage}}
\maketitle

\begin{abstract}

We apply probabilistic generative modelling of colour-magnitude diagrams to six young Galactic open star clusters and determine their mass functions, binary mass-ratio distributions, and the frequencies of binary stars. We find that younger clusters tend to exhibit a higher incidence of binaries than their older counterparts. 
The mass-ratio distribution is fairly flat for the clusters with one exception that exhibits a sharp increase for $q\gtrsim0.9$. The ratio of the number of cluster binaries for which $q>0.75$ to the number of binaries for which $q>0.5$ (referred to as $FQ_{75}$) ranges from $\sim0.4 - 0.8$. This metric increases with the binary-star frequency of a cluster, but declines with cluster age.
This may be due to non-ionizing 3-body dynamical processing of a primordial population of close binaries
with initial mass ratios, $q \simeq 1$.


\end{abstract}




\section{Introduction}

Binary star systems have an enhanced affect on the evolution of star clusters when compared to single star systems. For instance, \citet{Hur2005} developed an N-body model, showing that a primordial binary star population is necessary to reproduce the observed number of blue stragglers in M67\citep{ahumada1995}. Gravitational interactions involving binaries are three-body or four-body interactions, which impart more energy to a cluster dynamical system than encounters between two single stars. Binaries act as an energy source in the cluster's core and influence the overall dynamical evolution. An increased initial binary fraction causes the gravitational contraction of a cluster to become more gradual and the core collapse more shallow \citep{Bhat_2023}.

Previous studies have determined the number of binaries in a star cluster using radial velocities, time series analysis, and colour-magnitude diagrams (CMDs). All techniques have their own biases and detection sensitivities. For example,
radial velocity detection becomes increasingly less sensitive as the orbital inclination or mass-ratio of a binary system decrease.
Time series analysis has been used to detect variable stars using statistical methods applied to periodic  light level changes \citep{albrow2001, Koentime} from, for example, detached and contact eclipsing binaries and BY Dra stars, however the inference of cluster binary frequency from such studies relies on assumptions, such as the magnetic braking timescale, that may not be robust.

In this paper, we focus on detection of binaries from Gaia CMDs.
Given the the colour and magnitude of a binary system are composite values of the individual stars, CMDs can distinguish between single stars and binaries. Generally binaries are
located above and red-wards 
of the single-star main sequence. Binaries with a low mass-ratio are difficult to distinguish from single stars, so it is standard to choose a threshold such as $q\geq0.5$ to quote the binary frequency. Previous determinations of the frequency of binary stars in Galactic open clusters include \citet{Li2020, Cordoni2023, Donada_2023, Yalyalieva2022, Yalyalieva2024}.

The mass-ratio distribution has been described by different shapes in various papers. In \cite{Duq91} the mass-ratio distribution was represented by a right skewed Gaussian with a peak at about $q\simeq0.23$ as it shows a trend towards small secondary masses, as also found by \cite{Kroupa90}. From a sample of 454 solar-type stars, \cite{Raghav2010} found a mass-ratio distribution that was  roughly uniform for $0.2\leq q\leq0.95$, with a deficiency of low-mass companions and a preference to like-mass pairs. \cite{Moe_2017} introduced three independent parameters to model the mass-ratio probability distribution. They used an initial power law for small mass-ratios ($0.1\leq q\leq0.3$) then a second power law slope for higher mass-ratios ($0.3\leq q\leq1.0$) and an excess fraction for $q>0.95$ relative to the previous power law. These results are generally incompatible with the approach suggested by \cite{Tout91}, where the secondary star in a binary system is taken randomly from the same initial mass function (IMF) as the primary (i.e. random pairings). The distributions could be more complicated however if the binary systems have undergone 
exchange transfers or mass transfer through Roche-lobe overflow \citep{Li_2022}.

In \cite{Moe_2017}, binary observations were systematically compiled to estimate the binary fraction across different spectral types. 40\% of solar-type stars were found to be a part of a binary or higher-order system. This fraction increased for A-, B-, and O-type stars. In the context of open clusters, the binary fraction ranges between 35-70\% \citep{Sollima_2010}. Recent findings by \citet{Donada_2023} revealed the median multiplicity fraction of 202 open clusters to be 18\% for $f_B(q>0.6)$. Binary systems are significantly rarer in globular clusters.  \cite{Milone_2012} found that the globular clusters had a significantly lower binary fraction than what was seen in the field. The binary fraction for these clusters could be as low as 5\%. 

This paper focuses on six young clusters located in the Milky Way. Collinder 69, $\alpha$ Persei, Pleiades, NGC 6405, Trumpler 10, and UPK 640. 
These were chosen from the Cantat-Gaudin+ catalogue \citep{Cantat} as being young (age $<10^8$ years) and populous ($>400$ members). The positions and mean proper motions of the six chosen clusters are given in Table~\ref{Position Table}.


\begin{figure}
    \centering
    \includegraphics[width=0.9\columnwidth]{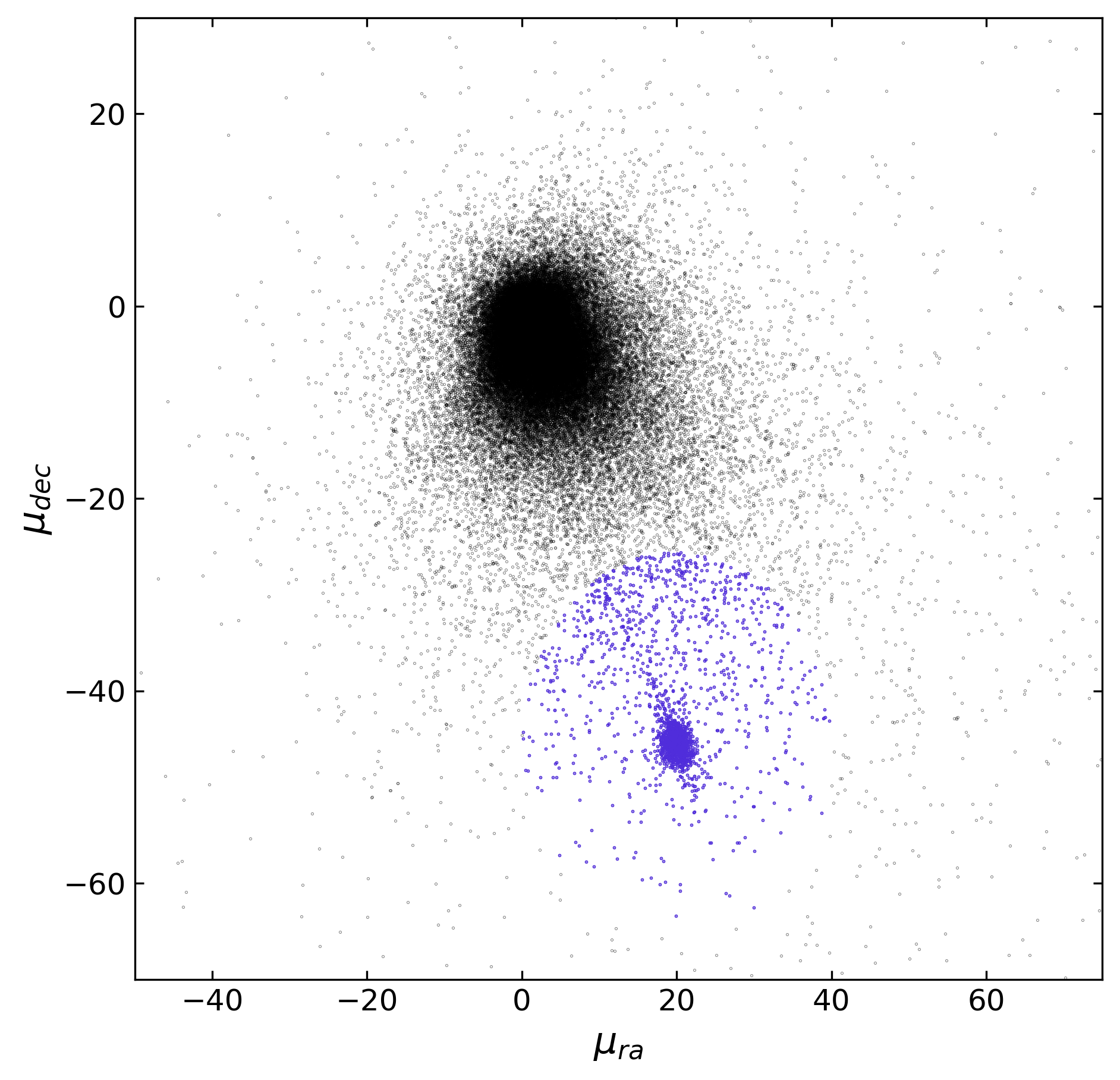}
    \caption{Proper motions of all stars within RA and declination selection around the centre of Pleiades. The black points are the background stars and the blue points are the stars included in our initial selection of possible cluster members.}
    \label{fig2:PM}
\end{figure}

\subsection{Collinder 69}

Also known as the Lambda Orionis cluster, Collinder 69 is a young cluster with an age between 5 and 16 million years \citep{Bayo_2008}. It lies within the Lambda Orionis star forming region where high-mass and low-mass stars began to form around a similar time until approximately 1 million years ago when this was abruptly terminated. Depending on the stellar model this was due to either the removal of the gas supply or births via cloud compression \citep{Dolan_1999}.

\subsection{$\alpha$ Persei}

$\alpha$ Persei (Melotte 20, Collinder 39) is a young nearby cluster of about 50 million years \citep{Mermilliod81} and contains a tight main-sequence in the CMD \citep{mermilliod2008}. It is
located in the vicinity of a thick stellar stream ($\mathrm{\sim180 pc}$ in the line of view) and the cluster has tidal tail structures that are likely of Galactic origin \citep{Nikiforova_2020}.

\subsection{Pleiades}

This is one of the brightest and most visible star clusters in the night sky hence it has many alternate names such as  Melotte 22, Messier 45, the Seven Sisters, Subaru, or Matariki. Pleiades is the oldest and largest of this selection of open clusters in this paper and has an age of 75 to 150 million years \citep{Dahm2015}. Due to its close proximity, it has been heavily studied by various authors to understand more general properties of star clusters \citep{Elsanhoury15}.

\subsection{NGC 6405}

Also referred to as either Messier 6 or the Butterfly Cluster, NGC 6405 is an open cluster with an age between 50 and 100 million years \citep{Lindoff68,paunzen2006}. This cluster is an interarm object lying between the local and Sagittarius arms \citep{Vleeming74} with a metallicity close to Solar. 

\subsection{Trumpler 10}

Recently $\delta$ Scuti stars were used to date Trumpler 10 to be between 10 and 60 million years \citep{Ortega23}. This cluster can be found in the Vela constellation and at the edge of the Gum Nebula where it could be partially responsible for the ionising fluxes \citep{kim2005}.

\subsection{UPK 640}

This is the second youngest cluster from our selection with an age of $\sim25-35$ million years \citep{Cantat,Sim2019} and is also referred to as LP 2442. This cluster has five hierarchical groups which form a filament structure and these groups are spatially well-mixed but kinematically coherent. It has been suggested that there could be an ongoing merging process between these subgroups \citep{Pang_2022}.

\begin{figure*}
     \centering
     \begin{subfigure}[b]{0.33\textwidth}
         \centering
         \includegraphics[width=\textwidth]{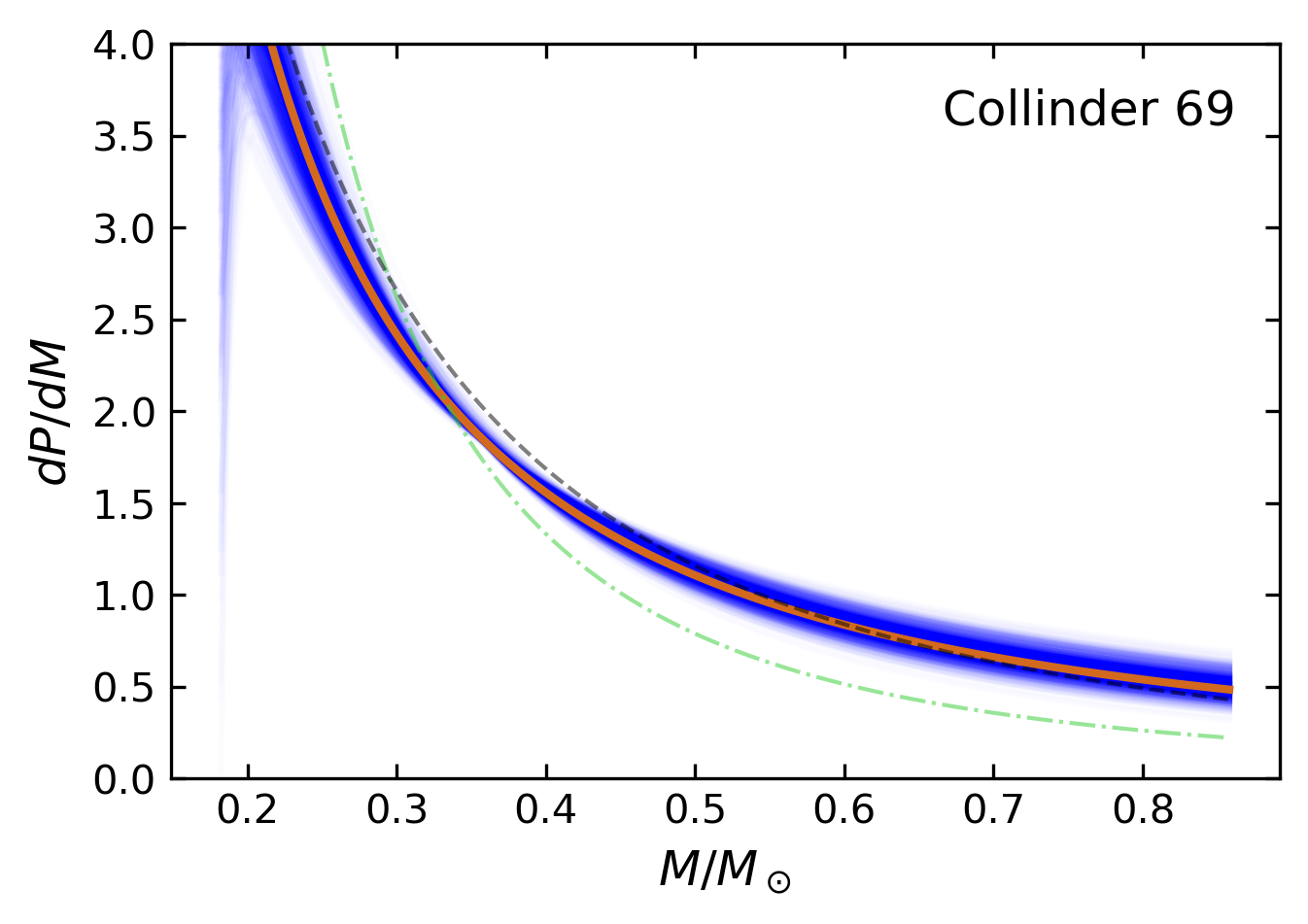}
         \label{C69 mf}
     \end{subfigure}
     \hfill
     \begin{subfigure}[b]{0.33\textwidth}
         \centering
         \includegraphics[width=\textwidth]{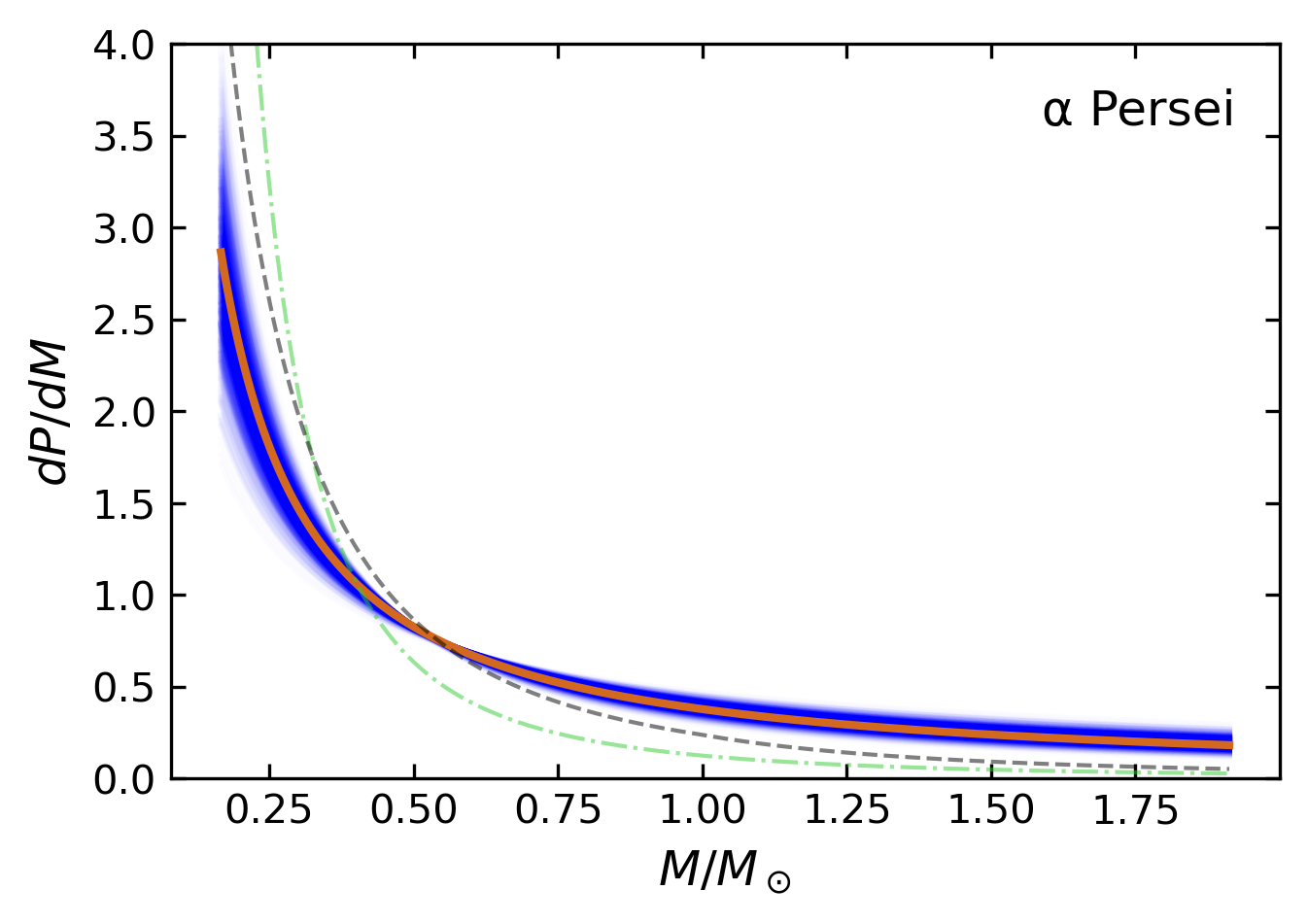}
         \label{Mel20 mf}
     \end{subfigure}
     \hfill
     \begin{subfigure}[b]{0.33\textwidth}
         \centering
         \includegraphics[width=\textwidth]{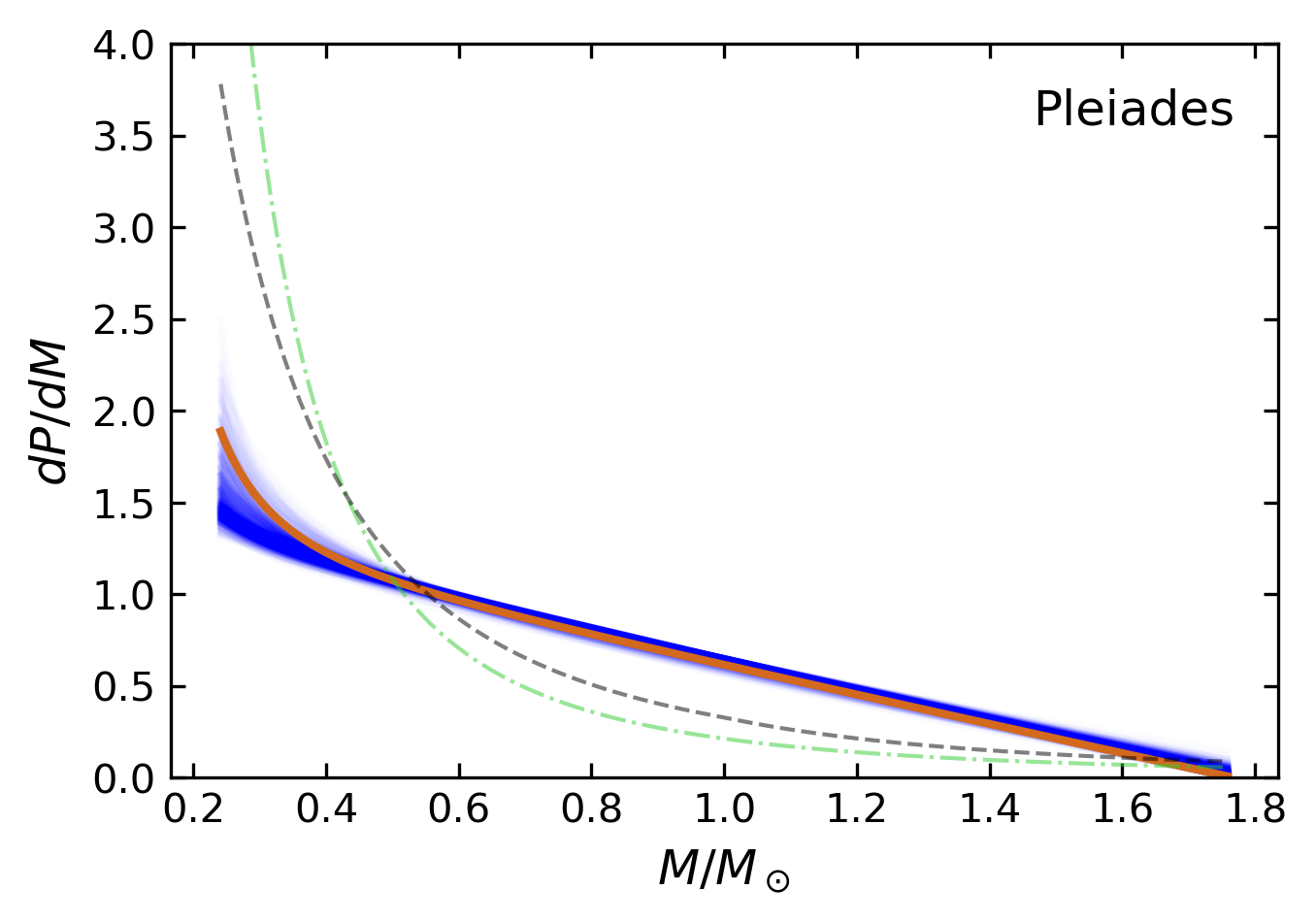}
         \label{Mel22 qdist}
     \end{subfigure}
        \centering
     \begin{subfigure}[b]{0.33\textwidth}
         \centering
         \includegraphics[width=\textwidth]{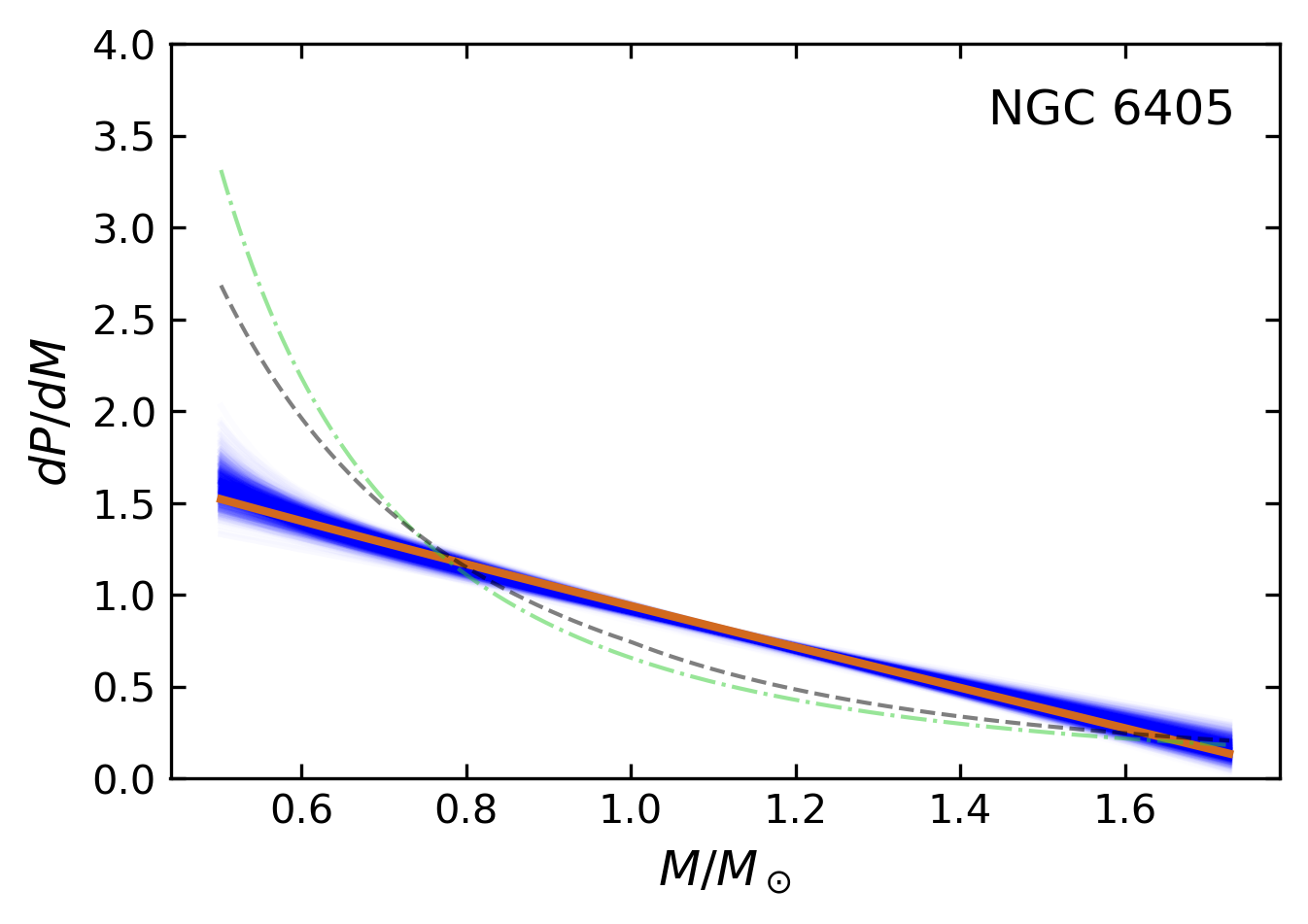}
         \label{NGC6405 mf}
     \end{subfigure}
     \hfill
     \begin{subfigure}[b]{0.33\textwidth}
         \centering
         \includegraphics[width=\textwidth]{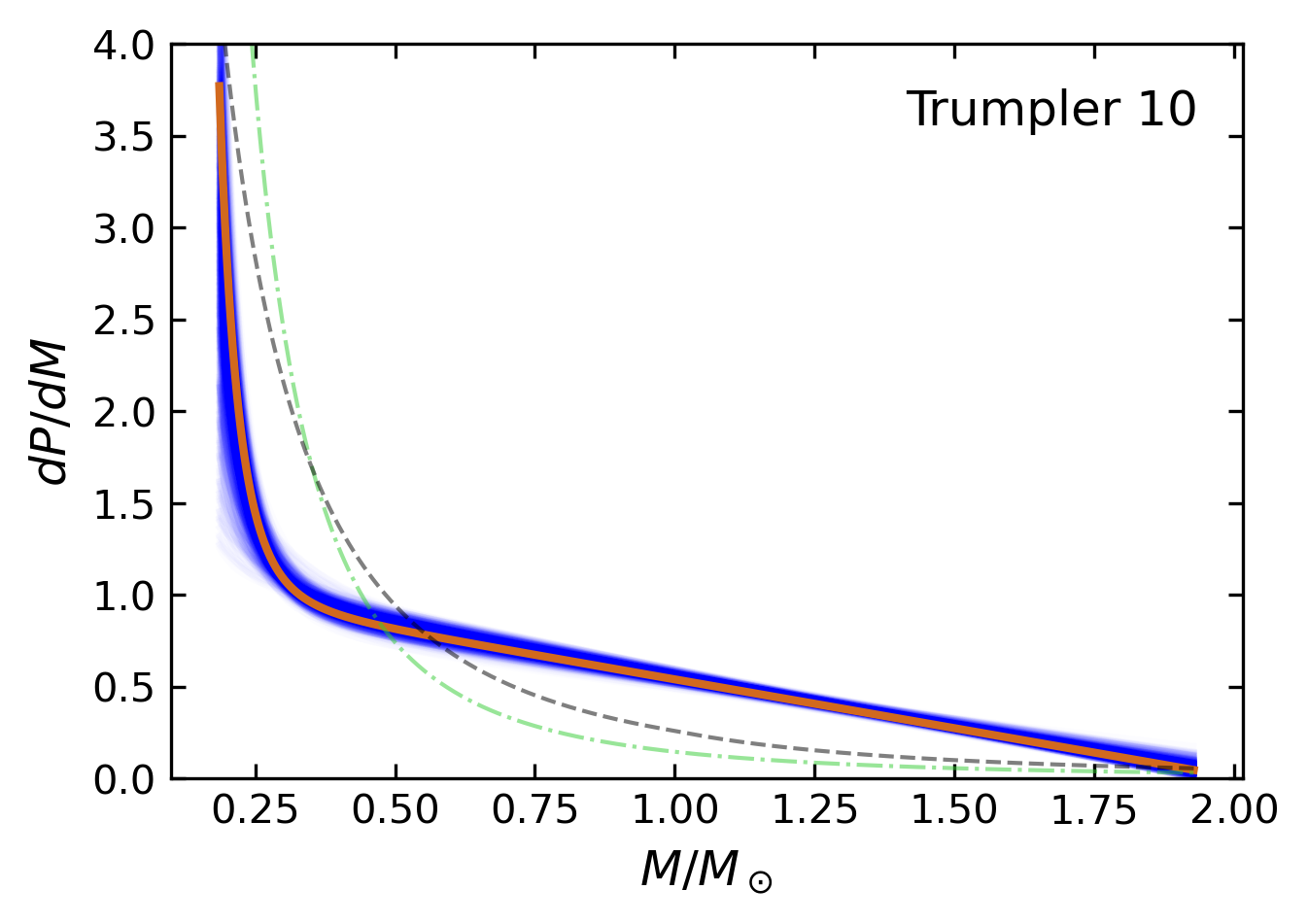}
         \label{TMP10 mf}
     \end{subfigure}
     \hfill
     \begin{subfigure}[b]{0.33\textwidth}
         \centering
         \includegraphics[width=\textwidth]{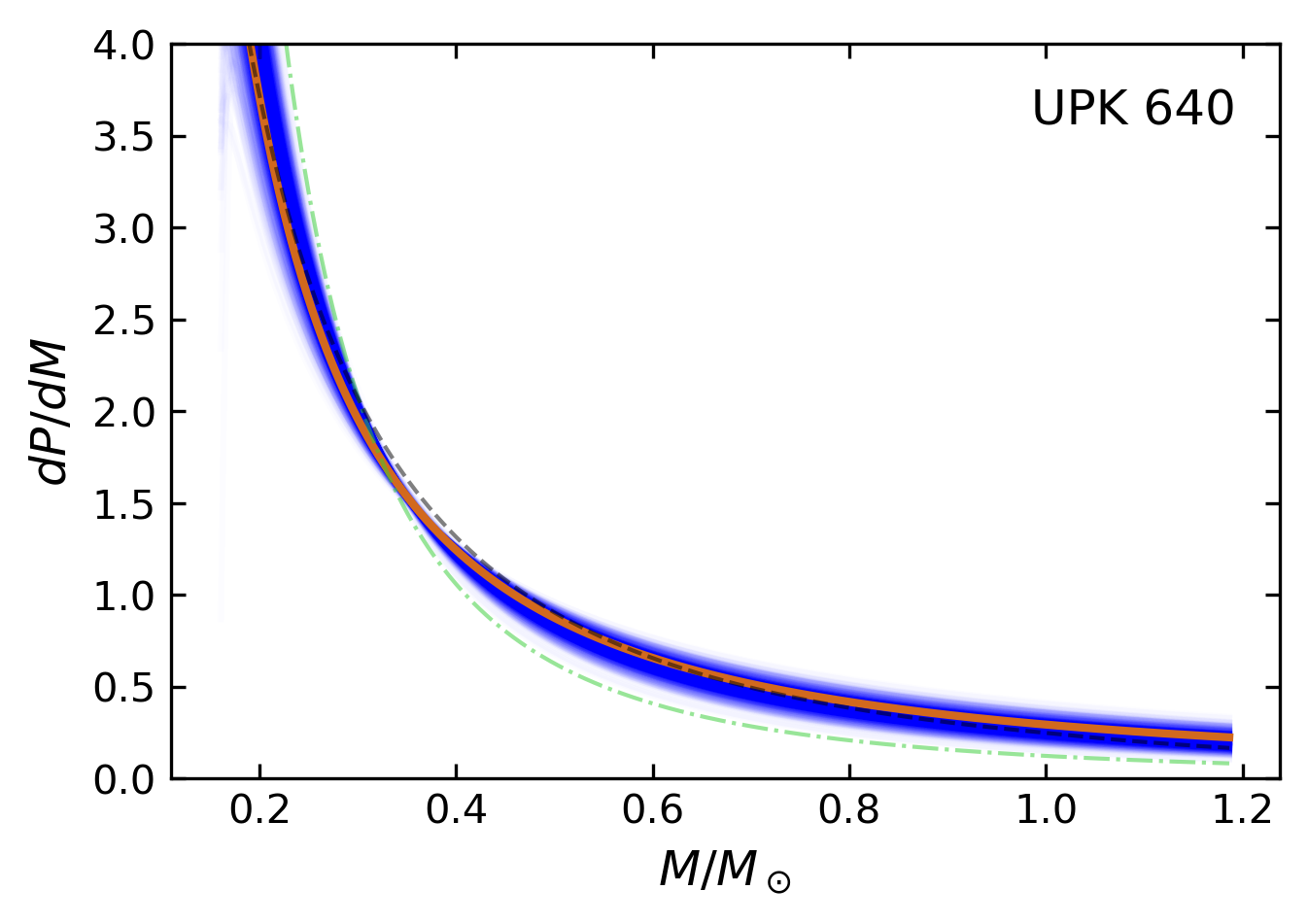}
         \label{UPK640 mf}
     \end{subfigure}
\caption{Mass functions for each cluster. Blue lines are based on 1000 random samples from the posterior distribution. The orange line shows the maximum probability solution. The \citet{Salpeter55} and \citet{Chabrier_2003} initial mass functions are shown as green and black dashed lines respectively.}
    \label{mf graphs}
\end{figure*}


\section{Data Selection and Filtering}

\subsection{Proper-motion and parallax}

Using data from Gaia DR3 we made a selection of all stars in a $4\times4$ degree square centred at the catalogued position of each cluster. From this data we used two methods to identify cluster-member stars.
A hard circular cut in proper motion space was applied based on the tabulated proper motion of each cluster 
(Table~\ref{Position Table}), with a radius approximately twice the range of stellar proper motions found in the catalogue. An example of this proper-motion cut is shown
in Fig.~\ref{fig2:PM}. We note that many of our clusters exhibit dynamic (proper motion) trails, and our selection does not necessarily include all members of a given cluster.

\begin{table}
    \centering
    \caption{Central positions and proper motions used for the initial selection of cluster members. $\umu_{Rad}$ denotes the radius of selection used.}
    \label{Position Table}
    \begin{tabular}{lcccccc}
         \hline
         Name & \small{RA}  & \small{dec}  & \small{$\umu_{\mathrm{RA}}$}  & \small{$\umu_{\mathrm{dec}}$} & $\umu_{\mathrm{Rad}}$ \\\
          & \scriptsize{deg} & \scriptsize{deg} & \scriptsize{mas yr$^{-1}$} & \scriptsize{mas yr$^{-1}$} & \scriptsize{mas yr$^{-1}$} \\
         \hline\hline
         Collinder 69 & 83.792 & 9.813 & 1.194 & -2.118 & 2\\
         $\alpha$ Persei & 51.617 & 48.975 & 22.864 & -25.378 & 10\\
         Pleiades & 56.601 & 24.114 &  20.077 & -45.503 & 20\\
         NGC 6405 & 265.069 & -32.242 & -1.306 & -5.847 & 2\\
         Trumpler 10 & 131.943 & -42.566 & -12.532 & 6.527 & 2\\
         UPK 640 & 250.260 & -39.494 & -11.973 & -21.250 & 10\\
         \hline
    \end{tabular}
\end{table}

We then removed stars with greater than a 30\% uncertainty in Gaia  parallax. A distance-selection was made by applying 
cuts either side of the peak in parallax that corresponded to the tabulated cluster distance.

Finally, we modelled the density of the remaining stars in proper-motion and parallax space with a three-dimensional Gaussian, 
rejecting those outside a probability threshold of 0.7.

During this selection process we followed the differential reddening procedure of \cite{Milone_2012}, however the impact was minimal and only resulted in more scatter. This was therefore omitted from the final data.

\subsection{DBSCAN}

As a check on the proper-motion and parallax selection method, we also applied the DBSCAN algorithm
to the original $4\times4$ degree$^2$ spatial data selection for each target. 
DBSCAN works by generating a hypersphere with a normalised radius of $\epsilon$ around each point and across each dimension 
(in this case RA, dec, $\mu_{\mathrm{RA}}$, $\mu_{\mathrm{dec}}$, and parallax) and uses {\em density reachability} to define clusters. Two points ${\bm x}$ and 
${\bm y}$ are directly density reachable if, for a distance metric $d$,
\begin{equation}
    d({\bm x},{\bm y})\leq\epsilon,
\end{equation}
and the two points are density reachable if there exists a chain of points; ${\bm p_i}$, between ${\bm x}$ and ${\bm y}$ such that 
\begin{equation}
    d({\bm p_{i-1}},{\bm p_i})\leq\epsilon,\quad d({\bm p_i},{\bm p_{i+1}})\leq\epsilon
\end{equation}
holds for every point. Clusters are classified by the collection of all density reachable points. 

We first normalised all data dimensions to have a mean of zero and standard deviation of one,
then used the Euclidean metric with the
maximum distance between points; $\epsilon$, set to be $2.0$ and the minimum number of points within each 4-sphere to be considered a core point set to 15 points. All points that did not meet both of these conditions were removed. 

In all cases, the final selection of data points was almost identical to the proper-motion and parallax selection made above.

\begin{table} 
    \centering
    \begin{threeparttable}
        \caption{Best-fitting log age and metallicity  used to create isochrones of each star cluster.}
        \label{Age & Metal}
        \begin{tabular}{lccc}
             \hline
             Name & Age & [Fe/H] & References\\
             &\scriptsize{$\log$ yr} & \scriptsize{dex} & \\
             \hline\hline
             Collinder 69 & 7.0 & -0.25 & [1], [2], [3]\\
             $\alpha$ Persei & 7.8 & 0.00 & [4], [2], [3]\\ 
             Pleiades & 7.9 & 0.00 & [5], [6]\\
             NGC 6405 & 7.9 & 0.00 & [7]\\
             Trumpler 10 & 7.7 & 0.00 & [7], [8]\\
             UPK 640 & 7.3 & 0.00 & [9]\\
             \hline
        \end{tabular}
        \begin{tablenotes}
                \small
                \item \quad$^1$\citet{Bayo_2008}, 
                $^2$\citet{Abdurrouf2022}, \item \quad$^3$\citet{Healy_2023}, $^4$\citet{Basri_1999}, \item \quad$^5$\citet{Dahm2015},  $^6$\citet{Soderblom_2009}, \item \quad$^7$\citet{Jadhav_2021}, \item \quad$^8$\citet{Ortega23}, \item \quad$^9$\citet{fürnkranz2023}
        \end{tablenotes}

    \end{threeparttable}
\end{table}

\section{Model}

We adopt a modified version of the probabilistic generative model for colour-magnitude diagrams as detailed in \citet{Albrow22Ulusele} and \citet{Albrow23}. Each cluster is represented by the data set $\textbf{D}=\{\textbf{D}_k\}$ 
for stars $k$, occupying locations  ${\bm D_k} = (G-G_{RP},M_{G})_k^T$ on the CMD with
uncertainty covariance $\mathbfss{S}_k$.
These associated covariant uncertainties may be scaled such that $\mathbfss{S}_k \xrightarrow{} h^2 \mathbfss{S}_k$, where we define $h=h_0+\dot{h}(G_k-G_0)$ with $G_0$ as the central value of the photometric data, and $h_0$ and $\dot{h}$ are parameters of the model. 

\subsection{Isochrones}

For this study we used isochrones from the  MESA Isochrones and Stellar Tracks (MIST - \cite{MIST1} project \cite{MIST2}, \& \cite{MIST3}). The ages and metallicities of the isochrones used for each cluster are found in Table \ref{Age & Metal} along with the relevant references. These correspond to the isochrones which we found to best fit our results from the Section 2. MESA isochrones have been found to under-predict the stellar luminosity of single stars with masses between $0.25 M_\odot\leq M\leq0.85 M_\odot$ \citep{Brandner2023}. To account for this discrepancy, we adjust the $M_G$ magnitude of our isochrones to match the MS ridge line.

\subsection{Likelihood Function}

The likelihood function for a single data point is given by a mixture model, 
\begin{align}
    P(\textbf{D}_k|\bm{\theta})=&(1-f_B-f_O)P_S(\textbf{D}_k|\bm{\theta})+f_BP_B(\textbf{D}_k|\bm{\theta})\\&\nonumber+f_OP_O(\textbf{D}_k|\bm{\theta}),
\end{align}
where the likelihood functions for single stars, binary stars, and outliers are given by $P_S(\textbf{D}_k|\bm{\theta})$, $P_B(\textbf{D}_k|\bm{\theta})$, and $P_O(\textbf{D}_k|\bm{\theta})$ respectively as defined in \citet{Albrow22Ulusele, Albrow23}. The fraction of binaries is $f_B$ and the fraction of outliers is $f_O$, hence the fraction of single stars is $(1-f_B-f_O)$. 

We assume that single stars are positioned along an isochrone according to a parameterised mass distribution. We also assume for binaries that the more massive star is drawn from the same mass distribution as the single stars and that the mass ratio, $q$ is drawn from its own parameterised probability distribution. Both parameterised functions are decomposed into linear combinations of Gaussian basis functions. Using the isochrone, a pair of these basis functions in $M$ and $q$ maps onto a bivariate Gaussian on the colour-magnitude diagram. 

\subsection{Mass Function}

We adopt the form of the mass function  from \cite{Albrow23},
\begin{equation}
\begin{split}
\frac{dP(M|\gamma,k,M_{\rm 0},c_0,c_1)}{dM} =  
C_M \times \left( [(c_0 + c_1(M-M_{\rm min})] s_c  + M^{-\gamma} \right) \\  
 \times H(M - M_{\rm min}) \times H(M_{\rm max}-M)\\ \times \tanh(-k(M-M_0)).
\end{split}
\end{equation}
This consists of a power law plus a scaled linear function, which is then modified by sharp mass cut-offs at the top and bottom of the 
main sequence (the Heaviside functions , $H$) and a $\tanh$ function to allow for loss of detection sensitivity towards the
bottom of the main sequence. The latter function can be effectively excluded by setting $k$ and $M_0$ to appropriate constants.

\subsection{Mass-Ratio Distribution Function}
\label{section:mrdf}
Again, we adopt the two forms used in \citet{Albrow23}. The first of these consists of a sum of six shifted Legendre polynomials,
 \begin{equation}
\frac{dP(q | p_1, ..., p_{N_q})}{dq} = \tilde{P}_0(q) + \Sigma_{k=1}^{5} a_k  \tilde{P}_k(q),
\end{equation}
with five model parameters $a_k$. These coefficients are drawn from  normal distributions, 
$a_i \sim \mathcal{N}(0, \sigma)$, where $\sigma$ is treated as a hyperparameter, drawn from a reasonably-uninformative gamma distribution distribution, $\Gamma(2,3)$. 
This model has been found to produce the smoothest and flattest analytic functional forms that are consistent with the data.
We also use a histogram representation, with 10 bins.

\subsection{Binary Fraction}

We allow the binary fraction $f_B$ to vary along the main sequence. This is realised  as $f_B(M)=f_{B,0}+\dot{f_{B}}(M_{\mathrm{max}}-M),$ with model parameters $f_{B,0}$ and $\dot{f_{B}}$. As q decreases it becomes increasingly more difficult to distinguish between single stars and binary stars so we typically quote our binary fraction as $f_B(q>0.5)$, integrated over the main sequence.

\subsection{Priors}

Our model uses 16 or 20 free parameters, 10 of which arise from the mass function, error scaling, binary fraction, and outlier fraction such that $\bm{\theta_1}=\{\gamma,c_0,c_1,M_0,k,h_0,\dot{h},f_{B,0},\dot{f_B},f_O\}$.  For these 10 parameters, we adopt sensible uniform priors.
The remaining free parameters accrue from the mass-ratio distribution function, either $a_1 ... a_5$ and $\sigma$ with priors discussed in
Section~\ref{section:mrdf}, or the histogram values with Dirichlet priors \citep{Albrow23}.

The probability distribution for the entire vector of parameters, $\bm{\theta}$, is given by Bayes theorem,
\begin{equation}
    P(\bm{\theta}|\textbf{D})=\dfrac{P(\textbf{D}|\bm{\theta})P(\bm{\theta})}{Z}, \label{Bayes}
\end{equation}
where $P(\textbf{D}|\bm{\theta})$ is the likelihood function, $P(\bm{\theta})$ are our parameter prior values, and $Z$ is the model evidence 
(marginal likelihood). The model code allows different parameters or parameter combinations to be set as constants (frozen), thus reducing the dimensionality of the model. To compare models with different numbers of parameters we use the evidence, $Z$, which guards against overparameterization by penalising models with more parameters.

\section{Results}

\begin{figure*}
     \centering
     \begin{subfigure}[b]{0.33\textwidth}
         \centering
         \includegraphics[width=\textwidth]{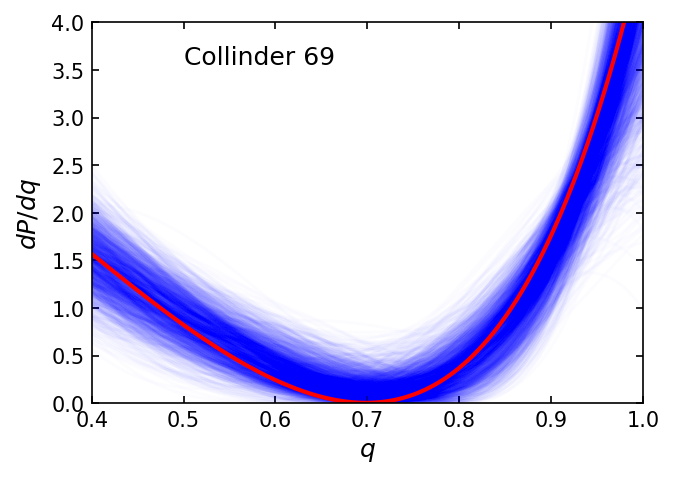}
         \label{C69 qdist}
     \end{subfigure}
     \hfill
     \begin{subfigure}[b]{0.33\textwidth}
         \centering
         \includegraphics[width=\textwidth]{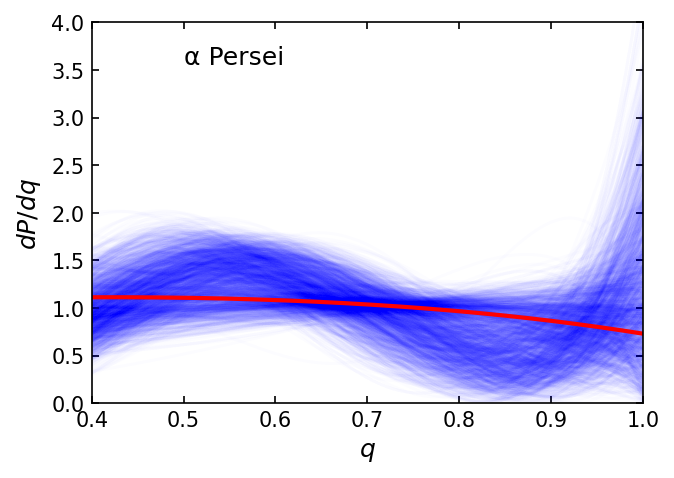}
         \label{Mel20 qdist}
     \end{subfigure}
     \hfill
     \begin{subfigure}[b]{0.33\textwidth}
         \centering
         \includegraphics[width=\textwidth]{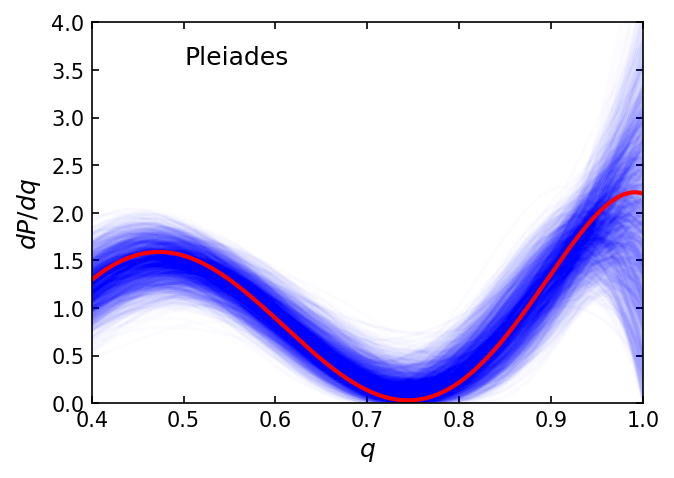}
         \label{Mel22 qdist}
     \end{subfigure}
        \centering
     \begin{subfigure}[b]{0.33\textwidth}
         \centering
         \includegraphics[width=\textwidth]{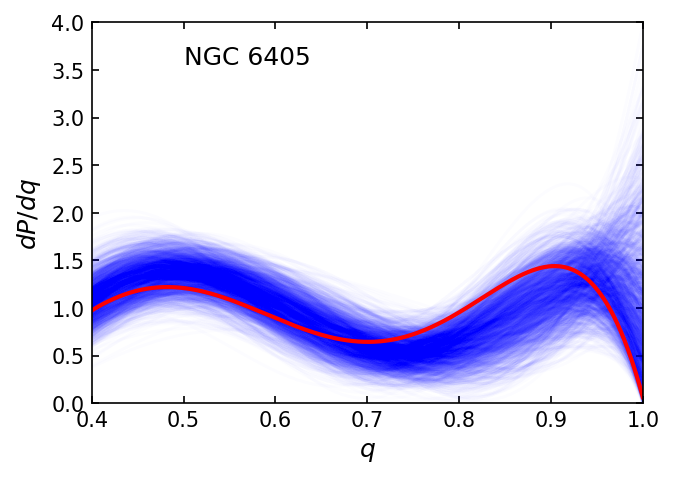}
         \label{NGC6405 qdist}
     \end{subfigure}
     \hfill
     \begin{subfigure}[b]{0.33\textwidth}
         \centering
         \includegraphics[width=\textwidth]{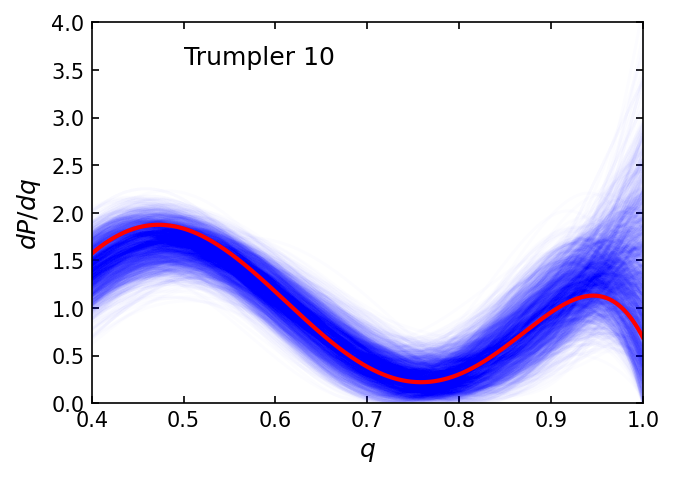}
         \label{TMP10 qdist}
     \end{subfigure}
     \hfill
     \begin{subfigure}[b]{0.33\textwidth}
         \centering
         \includegraphics[width=\textwidth]{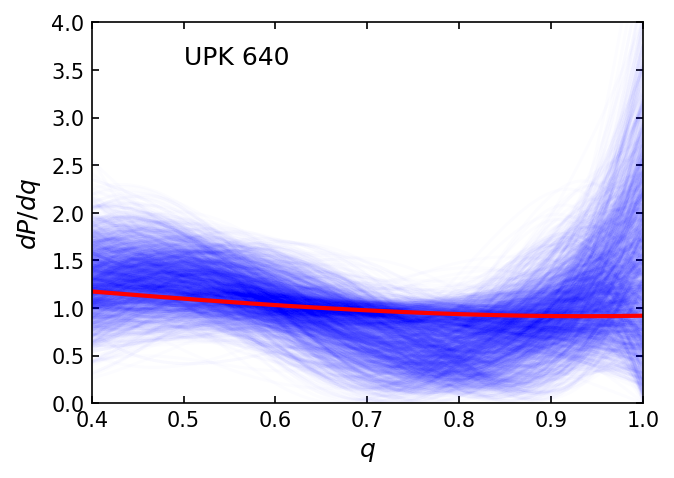}
         \label{UPK640 qdist}
     \end{subfigure}
     \begin{subfigure}[b]{0.33\textwidth}
         \centering
         \includegraphics[width=\textwidth]{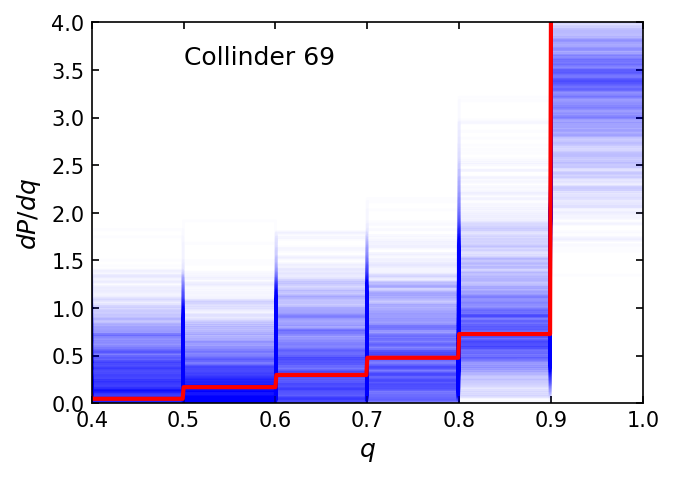}
         \label{HC69 qdist}
     \end{subfigure}
     \hfill
     \begin{subfigure}[b]{0.33\textwidth}
         \centering
         \includegraphics[width=\textwidth]{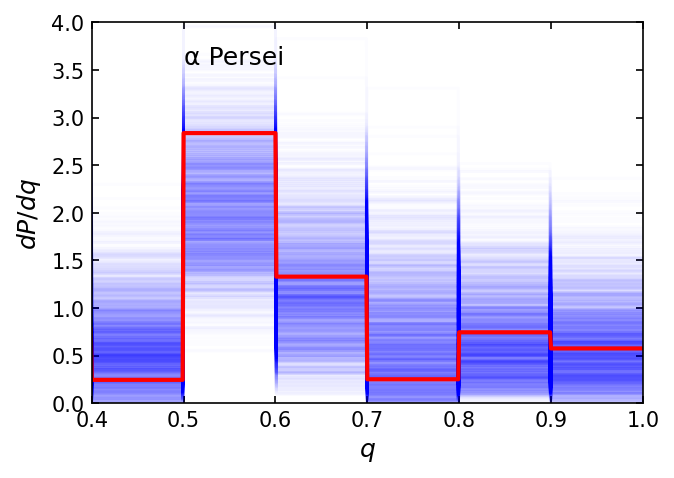}
         \label{HMel20 qdist}
     \end{subfigure}
     \hfill
     \begin{subfigure}[b]{0.33\textwidth}
         \centering
         \includegraphics[width=\textwidth]{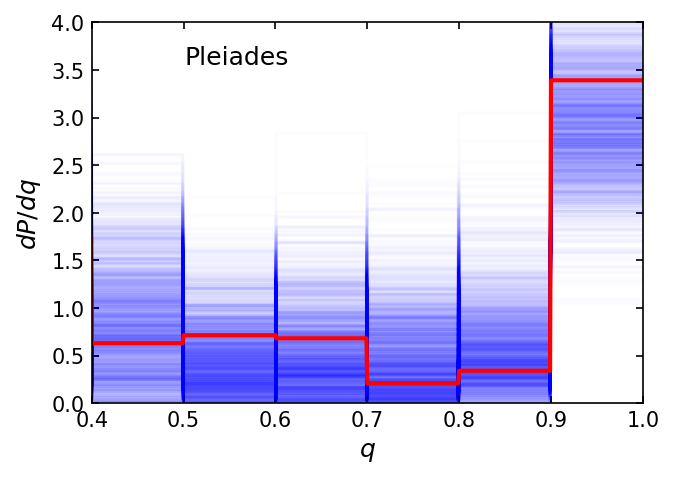}
         \label{HMel22 qdist}
     \end{subfigure}
        \centering
     \begin{subfigure}[b]{0.33\textwidth}
         \centering
         \includegraphics[width=\textwidth]{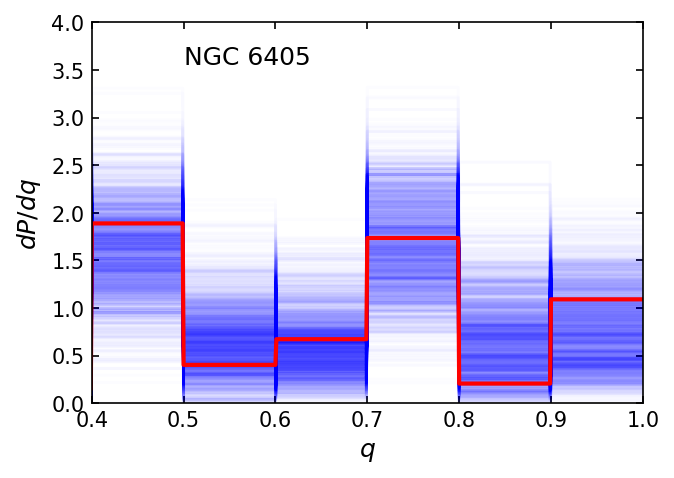}
         \label{HNGC6405 qdist}
     \end{subfigure}
     \hfill
     \begin{subfigure}[b]{0.33\textwidth}
         \centering
         \includegraphics[width=\textwidth]{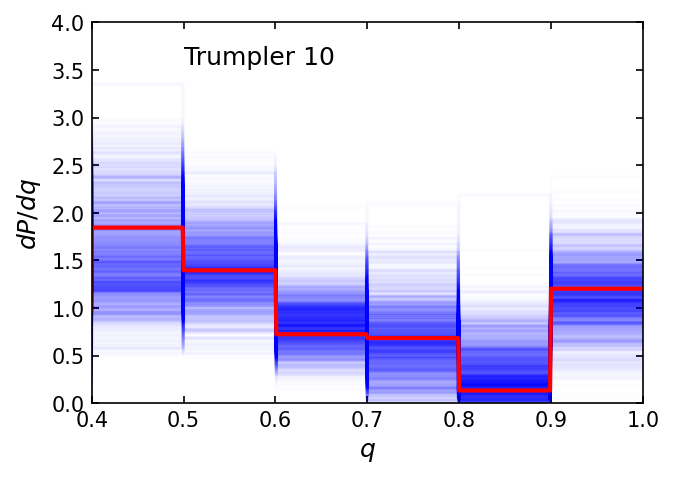}
         \label{HTMP10 qdist}
     \end{subfigure}
     \hfill
     \begin{subfigure}[b]{0.33\textwidth}
         \centering
         \includegraphics[width=\textwidth]{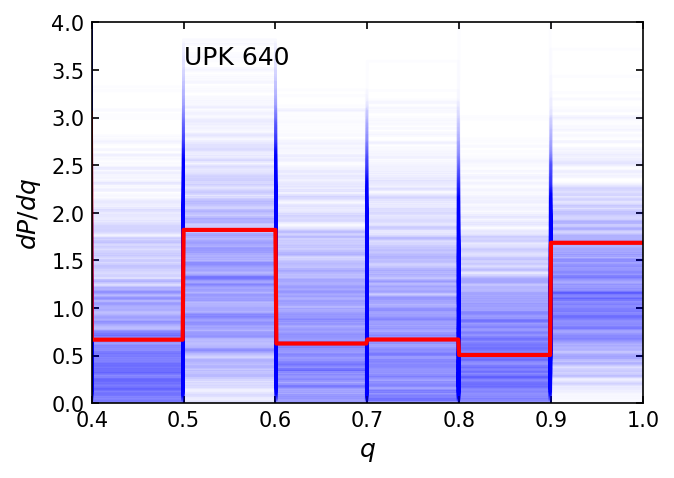}
         \label{HUPK640 qdist}
     \end{subfigure}
        \caption{The mass-ratio distribution using Legendre polynomials and histograms with ten bins for each cluster. The blue lines are 1000 random samples from the posterior distribution, and the red line indicates the maximum probability solution.}
        \label{qdist graphs}
\end{figure*}

\begin{table*}
    \centering
    \caption{Median fraction of binary stars with mass ratio greater than a given q, $FQ_{75}$ (which measures the ratio of binaries where $q>0.75$ compared to those where $\mathrm{q}>0.5$), number of stars considered, and mass range of primary stars.}
    \label{q values}
    \begin{tabular}{lcccccc}
         \hline
         & \multicolumn{2}{c}{Collinder 69}& \multicolumn{2}{c}{$\alpha$ Persei}& \multicolumn{2}{c}{Pleiades}\\
         &  Legendre&Histograms (10 Bins)&  Legendre&Histograms (10 Bins)&  Legendre&Histograms (10 Bins)\\
        \scriptsize{$\mathrm{q^\prime}$} & \scriptsize{$\mathrm{f_B(q\geq q^\prime)}$} & \scriptsize{$\mathrm{f_B(q\geq q^\prime)}$} & \scriptsize{$\mathrm{f_B(q\geq q^\prime)}$} & \scriptsize{$\mathrm{f_B(q\geq q^\prime)}$} & \scriptsize{$\mathrm{f_B(q\geq q^\prime)}$} & \scriptsize{$\mathrm{f_B(q\geq q^\prime)}$} \\
         \hline\hline
        0.4 & $0.341^{+0.036}_{-0.038}$ & $0.276^{+0.037}_{-0.038}$ & $0.262^{+0.052}_{-0.050}$  & $0.235^{+0.037}_{-0.041}$ & $0.229^{+0.024}_{-0.028}$  & $0.199^{+0.025}_{-0.025}$  \vspace{0.1cm}\\
        0.5 & $0.272^{+0.034}_{-0.036}$  & $0.258^{+0.035}_{-0.036}$  & $0.211^{+0.040}_{-0.039}$  & $0.209^{+0.035}_{-0.038}$  & $0.174^{+0.019}_{-0.022}$  & $0.145^{+0.020}_{-0.020}$  \vspace{0.1cm}\\
        0.6 & $0.235^{+0.034}_{-0.038}$ & $0.241^{+0.034}_{-0.035}$  & $0.152^{+0.027}_{-0.030}$  & $0.129^{+0.028}_{-0.031}$  & $0.127^{+0.017}_{-0.020}$  & $0.121^{+0.019}_{-0.020}$  \vspace{0.1cm}\\
        0.7 & $0.220^{+0.036}_{-0.039}$  & $0.224^{+0.033}_{-0.035}$  & $0.101^{+0.023}_{-0.026}$  & $0.080^{+0.024}_{-0.027}$  & $0.105^{+0.018}_{-0.020}$  & $0.091^{+0.017}_{-0.019}$  \vspace{0.1cm}\\
        0.8 & $0.208^{+0.034}_{-0.038}$  & $0.203^{+0.035}_{-0.037}$  & $0.065^{+0.020}_{-0.023}$  & $0.051^{+0.018}_{-0.023}$  & $0.098^{+0.017}_{-0.019}$  & $0.081^{+0.016}_{-0.019}$  \vspace{0.1cm}\\
        0.9 & $0.158^{+0.028}_{-0.030}$  & $0.169^{+0.037}_{-0.037}$  & $0.036^{+0.013}_{-0.018}$  & $0.022^{+0.013}_{-0.019}$  & $0.068^{+0.014}_{-0.016}$  & $0.058^{+0.019}_{-0.019}$  \vspace{0.1cm}\\
         \hline
         $FQ_{\mathrm{75}}$ & $0.802_{-0.089}^{+0.070}$  & $0.838_{-0.090}^{+0.073}$  & $0.398_{-0.111}^{+0.102}$   & $0.320_{-0.083}^{+0.092}$  & $0.590_{-0.080}^{+0.074}$  & $0.597_{-0.87}^{+0.089}$  \vspace{0.1cm}\\
         $N$ & \multicolumn{2}{c}{$297$}  & \multicolumn{2}{c}{$157$}  & \multicolumn{2}{c}{$323$}\\
         $M/M_\odot$ & \multicolumn{2}{c}{$[0.183,0.848]$}  & \multicolumn{2}{c}{$[0.171,1.733]$}  & \multicolumn{2}{c}{$[0.242,1.661]$}\\
         \hline
    \end{tabular}
    \begin{tabular}{lcccccc}
         \hline
         & \multicolumn{2}{c}{NGC 6405}& \multicolumn{2}{c}{Trumpler 10}& \multicolumn{2}{c}{UPK 640}\\
  &   Legendre&Histograms (10 Bins)&  Legendre&Histograms (10 Bins)& Legendre&Histograms (10 Bins)\\
 \scriptsize{$\mathrm{q^\prime}$} & \scriptsize{$\mathrm{f_B(q\geq q^\prime)}$} & \scriptsize{$\mathrm{f_B(q\geq q^\prime)}$} & \scriptsize{$\mathrm{f_B(q\geq q^\prime)}$} & \scriptsize{$\mathrm{f_B(q\geq q^\prime)}$} & \scriptsize{$\mathrm{f_B(q\geq q^\prime)}$} & \scriptsize{$\mathrm{f_B(q\geq q^\prime)}$} \\
         \hline\hline
        0.4 & $0.284^{+0.027}_{-0.028}$  & $0.263^{+0.029}_{-0.029}$ & $0.255^{+0.025}_{-0.027}$  & $0.225^{+0.025}_{-0.027}$ & $0.274^{+0.048}_{-0.050}$ & $0.249^{+0.040}_{-0.041}$ \vspace{0.1cm}\\
        0.5 & $0.223^{+0.021}_{-0.023}$  & $0.191^{+0.024}_{-0.028}$  & $0.186^{+0.020}_{-0.022}$  & $0.165^{+0.026}_{-0.025}$  & $0.214^{+0.039}_{-0.040}$ &  $0.226^{+0.038}_{-0.041}$  \vspace{0.1cm}\\
        0.6 & $0.163^{+0.019}_{-0.021}$  & $0.163^{+0.021}_{-0.023}$  & $0.124^{+0.018}_{-0.020}$  & $0.111^{+0.018}_{-0.019}$  & $0.158^{+0.031}_{-0.035}$ & $0.167^{+0.037}_{-0.040}$  \vspace{0.1cm}\\
        0.7 & $0.125^{+0.020}_{-0.023}$  & $0.136^{+0.021}_{-0.024}$  & $0.091^{+0.018}_{-0.020}$  & $0.080^{+0.015}_{-0.017}$  & $0.116^{+0.028}_{-0.034}$ & $0.116^{+0.031}_{-0.036}$  \vspace{0.1cm}\\
        0.8 & $0.099^{+0.019}_{-0.020}$  & $0.062^{+0.016}_{-0.019}$  & $0.078^{+0.018}_{-0.020}$  & $0.054^{+0.011}_{-0.015}$  & $0.084^{+0.027}_{-0.034}$ & $0.091^{+0.028}_{-0.031}$  \vspace{0.1cm}\\
        0.9 & $0.057^{+0.013}_{-0.015}$  & $0.032^{+0.015}_{-0.019}$  & $0.052^{+0.013}_{-0.015}$  & $0.043^{+0.012}_{-0.012}$  & $0.050^{+0.020}_{-0.028}$ & $0.068^{+0.028}_{-0.031}$  \vspace{0.1cm}\\
         \hline
         $FQ_{\mathrm{75}}$ &  $0.505_{-0.083}^{+0.080}$    & $0.519_{-0.063}^{+0.071}$  & $0.454_{-0.086}^{+0.090}$   & $0.408_{-0.058}^{+0.071}$  & $0.474_{-0.107}^{+0.106}$  & $0.474_{-0.119}^{+0.121}$  \vspace{0.1cm}\\
         $N$ & \multicolumn{2}{c}{$378$}  & \multicolumn{2}{c}{$290$}  & \multicolumn{2}{c}{$145$}\\
         $M/M_\odot$ & \multicolumn{2}{c}{$[0.505,1.727]$}  & \multicolumn{2}{c}{$[0.186,1.925]$}  & \multicolumn{2}{c}{$[0.161,1.139]$} \\
         \hline
    \end{tabular}
\end{table*}

\begin{figure*}
    \centering
    \includegraphics[width=\textwidth]{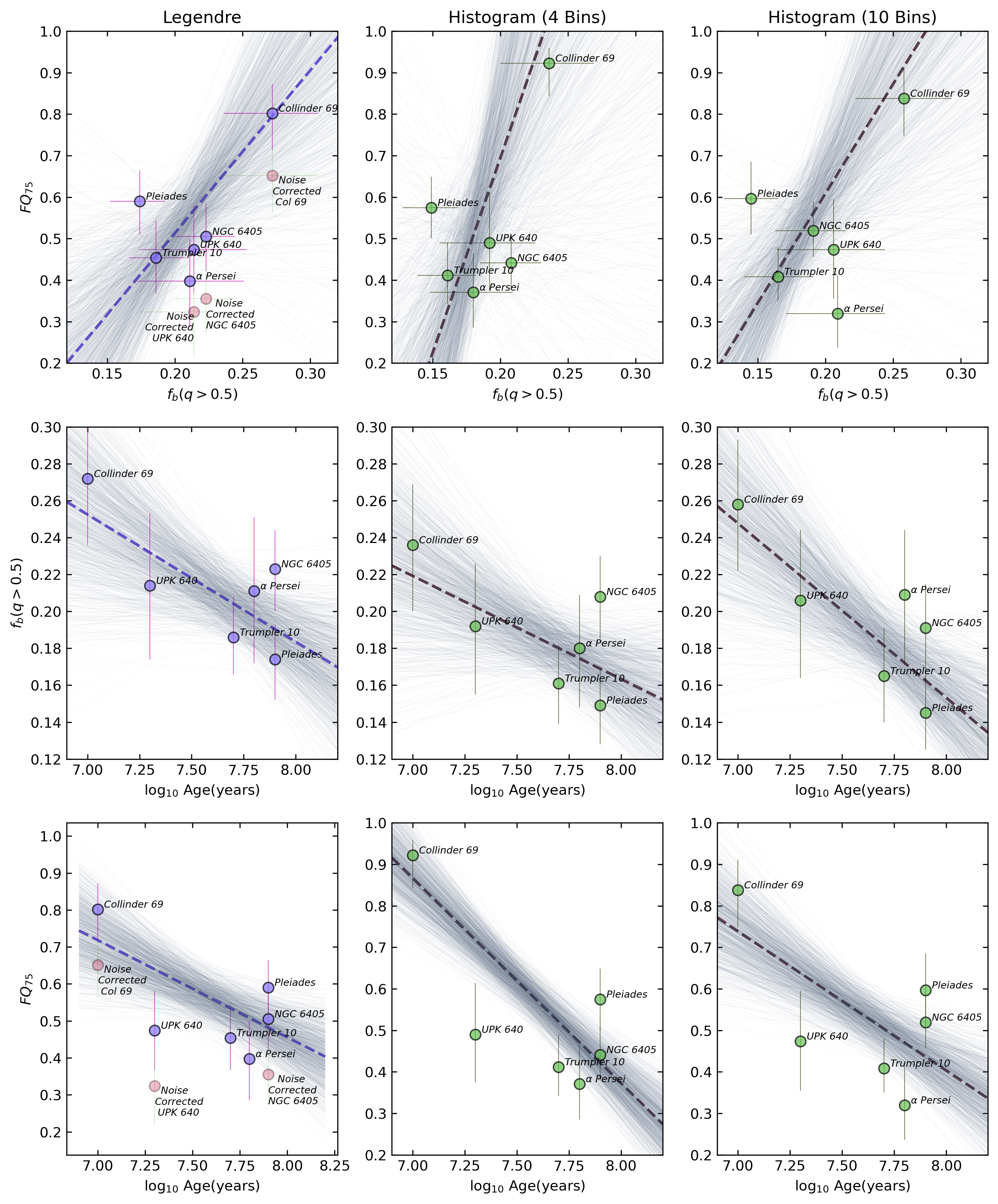}
    \caption{Inter-relations between $f_B, FQ_{75}$ and cluster age for different mass-ratio distribution function representations.}
    \label{LH4_10Bins}
\end{figure*}

For each of our clusters, the generalised form of nested sampling; dynamic nested sampling \citep{DynamicSampling}, was used to sample our posterior parameter distributions. For this we used the DYNESTY \citep{DYNESTY}  package which estimates the marginal likelihoods (evidence) and Bayesian posteriors. Our code allows for parameters to be frozen in various combinations, so the evidence was used to determine the best model to describe the data in each case. 

The evidence showed a preference for the binary fraction to vary along the main sequence ($\dot f_B\neq0$) for each cluster, apart from UPK 640 and NGC 6405. No cluster showed higher evidence when the uncertainty scaling was allowed to vary along the main sequence ($\dot h\neq0$). Both Collinder 69 and UPK 640 showed higher evidence when the mass function included a $\tanh$ function ($k\neq0,\;M_0\neq0$). $\alpha$ Persei, NGC 6405, and Trumpler 10 preferred a model of the mass function to include a scaled linear function alongside a power law, whereas the mass functions of the other clusters were best defined by only a power law ($c_0=0,\;c_1=0$). 

In the appendix we include a table of the parameters used by the best model for each cluster (Table \ref{Posterior params}) and the corner plots of the posteriors for each cluster (Figures \ref{fig:Col69 Corner}-\ref{fig:Upk640 Corner}). With the exception of 
$\alpha$ Persei and UPK 640, the solutions are unimodal. The solution spaces for
$\alpha$ Persei and UPK 640 are bimodal, with an extra solution that represents a 
maximal overall binary-star fraction, $f_B$. These solutions are effectively counting 
single main-sequence stars as low-mass-ratio binaries. This does not effect our results to follow, which are only derived for binary stars with $q > 0.5$.


\subsection{Mass Functions}
Figure \ref{mf graphs} shows the mass function for each of the six clusters. 
Generally, most of the clusters follow the expected behaviour of rising steeply towards lower masses, and are comparable to the canonical mass functions of \citep{Chabrier_2003} and \citep{Salpeter55}. The three clusters (Pleiades, NGC 6405, and Trumpler 10) that differ the most from the canonical distributions have flatter mass functions which include the scaled linear function. Note that the lowest mass stars from NGC 6405 are more massive than those from the other clusters, and the mass range can be found in Table \ref{q values}.



\subsection{Mass-ratio Distributions}
The mass-ratio distribution of each cluster, normalised over the range $0.4\leq q\leq1.0$, is displayed in Figure \ref{qdist graphs} 
for both the Legendre and histogram representations. We exclude $q < 0.4$, as single and binary stars are impossible to distinguish in these data for those low mass-ratios.
The shape of the distributions are consistent for all the clusters apart from Collinder 69. In Collinder 69, there is clearly a much higher frequency of high-mass-ratio binaries, whereas the other mass-ratio distributions are fairly flat. The mass-ratio distributions with the Legendre polynomial method typically show a minima from $0.6\leq q\leq0.8$, however this is not seen with the histogram representation. We believe this is likely a non-physical result which is being forced by the polynomial function.


\subsection{Realisations of the Data}
In the appendix (Fig.~\ref{Realisation} and Fig.~\ref{Realisation2}), we show colour-magnitude diagrams of the final result of our data selection alongside a randomly generated sample of three realisations for each of the clusters. Observational scatter has been provided by combining the median quoted data uncertainty as a function of magnitude for each cluster with the  median $h$
determined from the model. For a by-eye comparison of the data with the model realisations, the agreement is excellent.

\subsection{Binary Fraction and $FQ_{75}$}
The fraction of binaries above particular mass-ratios are given in Table \ref{q values} for both forms of the mass-ratio distribution function. These are produced by integrating the mass-ratio distribution from 1 to $q_{min}$. 

Collinder 69, the youngest cluster, has the highest binary fraction and about three times the fraction of high-mass-ratio ($0.9\leq q\leq1.0$) binaries than the other clusters. 

Also included in Table \ref{q values} is the metric $FQ_{75} \equiv f_B(q>0.75) / f_B(q>0.5) $ introduced in \cite{Albrow23}. This is a robust statistic designed to avoid the CMD degeneracy for binaries with $q \gtrsim 0.75$. Given the proclivity of high-mass-ratio binaries, it is unsurprising that $FQ_\mathrm{75}>0.5$ for Collinder 69 is significantly greater than the other clusters. $FQ_\mathrm{75}$ is greater than $0.5$ for Pleiades also, which reflects the  high frequency of high-mass-ratio binaries displayed in the mass-ratio distribution. For NGC 6405, $FQ_\mathrm{75}=0.505_{-0.083}^{+0.080}$, which correlates well with a flat mass-ratio distribution. For the remaining three clusters, $FQ_\mathrm{75}<0.5$ such that there is a lower proportion of high-mass-ratio binaries.



\subsection{Relations between $f_B, FQ_{75}$ and cluster age}
In Figure \ref{LH4_10Bins} we compare the binary fraction for $q\geq0.5$, $FQ_{75}$, and the age of each cluster for different models of the mass-ratio distribution. These models are defined by Legendre polynomials and histograms (as seen in \citet{Albrow23}), using 4 and 10 bins for the latter approach. 

We have made linear fits to these data using orthogonal distance regression. The best linear fit is plotted as a thick dashed line and we include the coefficients of these fits in Section \ref{Coeffcient Relationships} of the appendix. We generated 1000 samples of the linear fit from a multivariate distribution, and these have been plotted in grey. There is a positive correlation between the binary fraction and $FQ_\mathrm{75}$, a result consistent with the findings in \citet{Albrow23}. There is a negative correlation between the age of a cluster and the binary fraction, such that the frequency of binary systems is greater for the younger clusters. This is consistent with the trend found by \citet{Sollima_2010}. Transitive of the first two trends, there is also a negative correlation between the age of a cluster and $FQ_{75}$. 
From the generated samples we can see that there is some error to these fits. The positive correlation between the binary fraction and $FQ_\mathrm{75}$ is heavily dependent on the position of Collinder 69, such that there is the most variation between samples out of these relationships. The negative correlation between age and binary fraction also shows some variation, however the trend is fairly consistent. There tends to be a fairly strong agreement between samples for the negative correlation between the age of a cluster and $FQ_{75}$.



\subsection{Sensitivity to Isochrone Position}

\begin{figure*}
     \begin{subfigure}[b]{0.45\textwidth}
         \centering
    \includegraphics[width=\columnwidth]{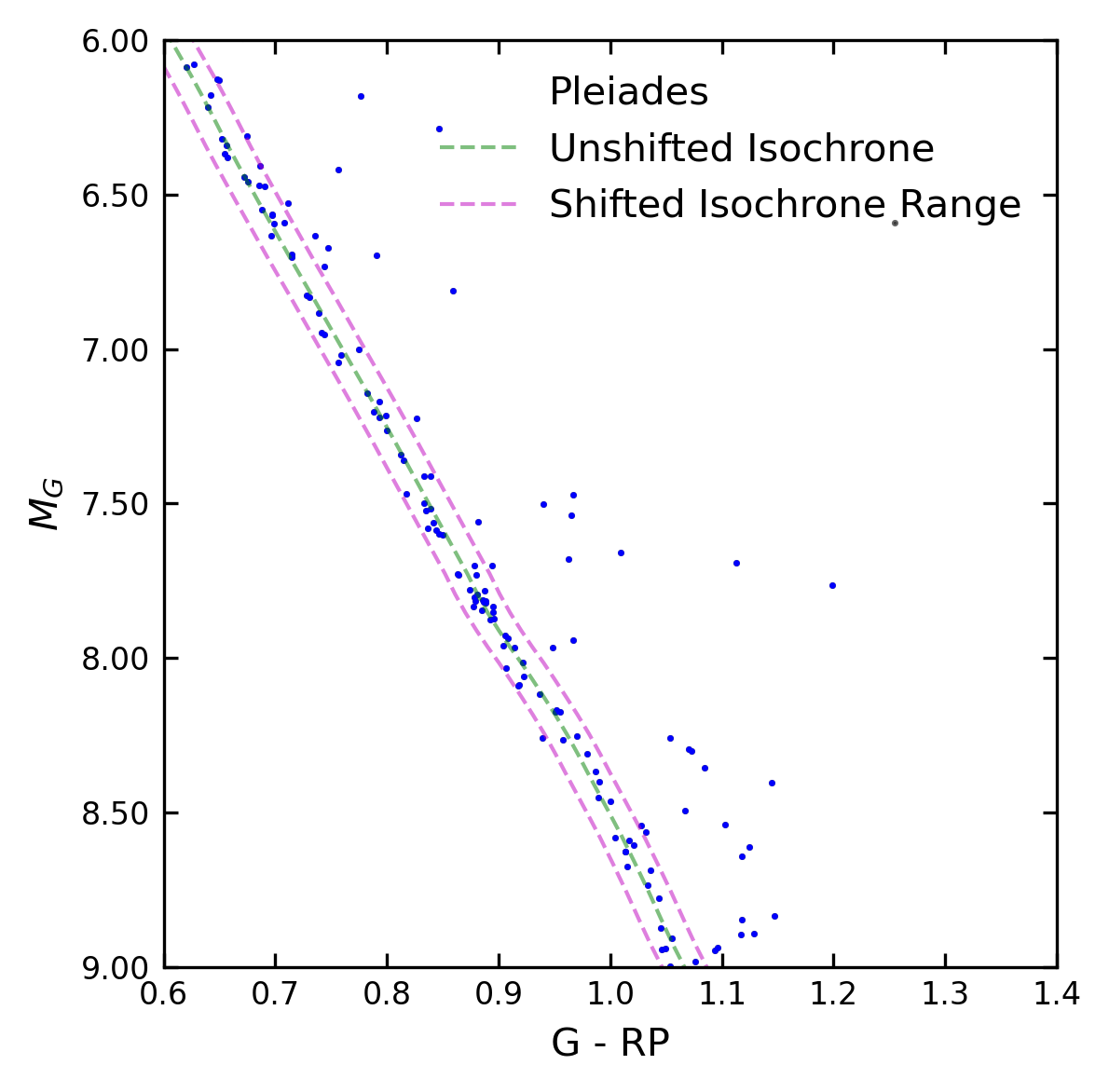}
     \end{subfigure}
      \hfill
     \begin{subfigure}[b]{0.45\textwidth}
         \centering
         \includegraphics[width=\columnwidth]{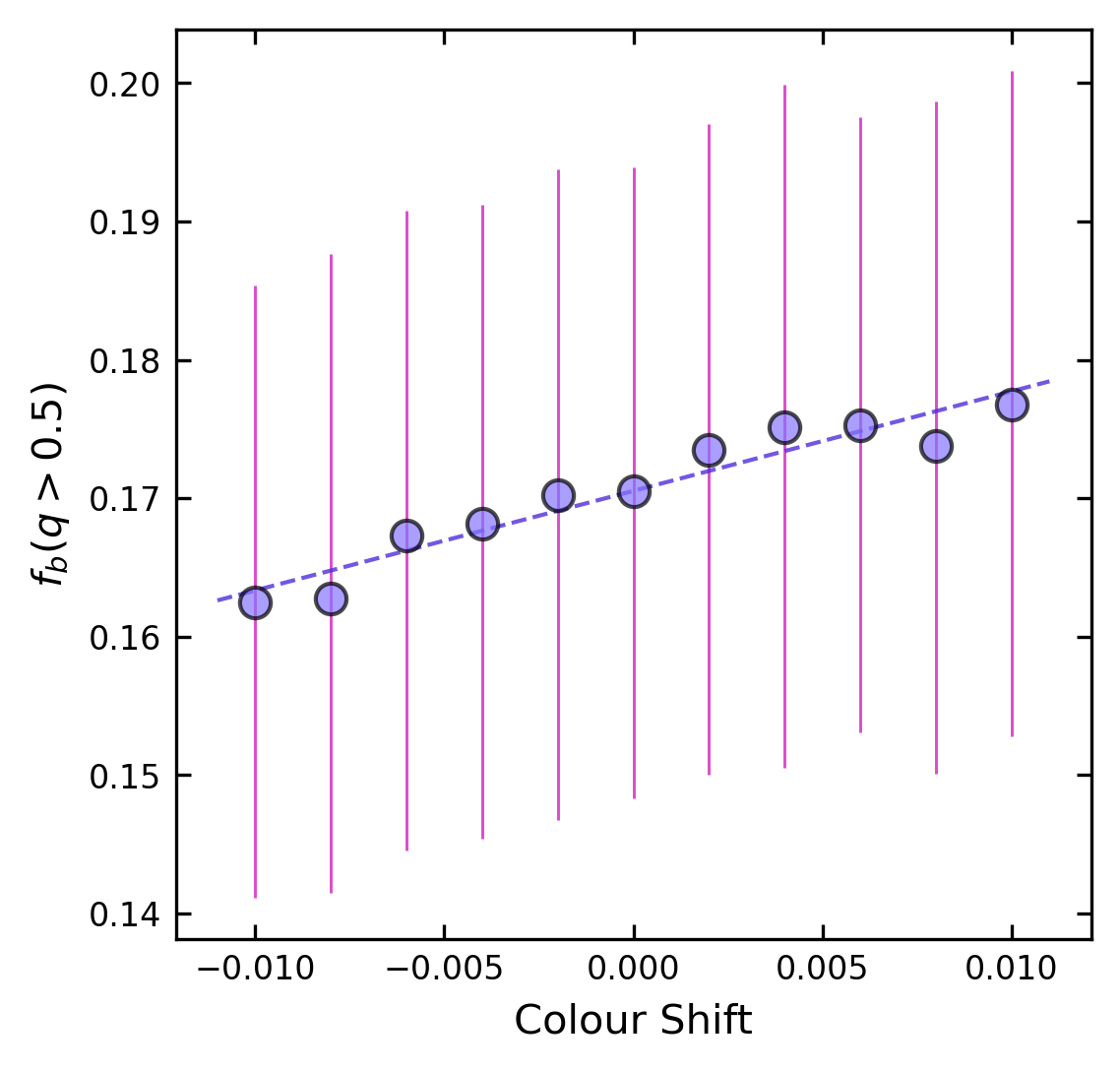}
     \end{subfigure}
      \hfill
     \begin{subfigure}[b]{0.45\textwidth}
         \centering
         \includegraphics[width=\columnwidth]{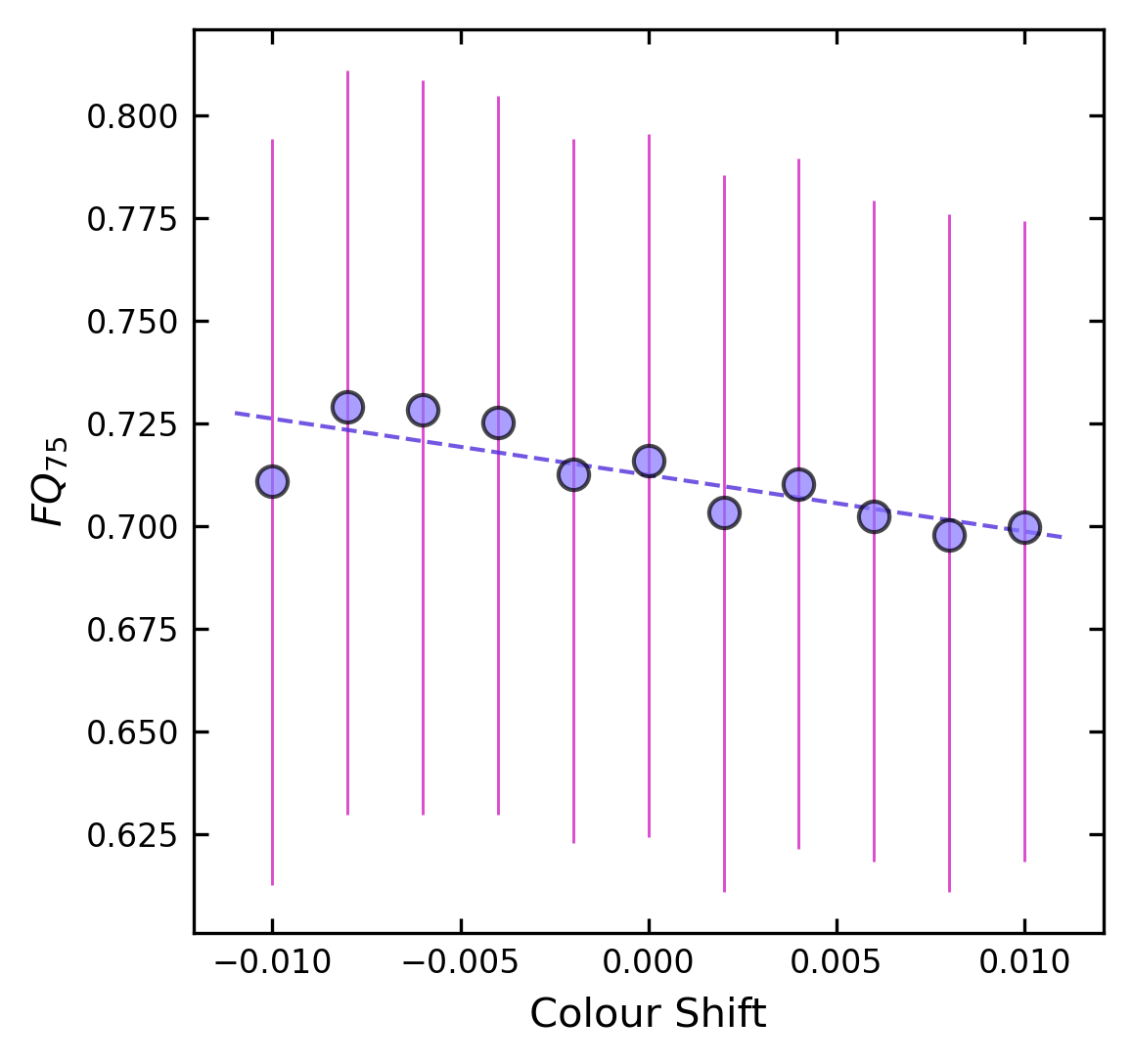}
     \end{subfigure}
     \hfill
     \begin{subfigure}[b]{0.45\textwidth}
         \centering
         \includegraphics[width=\columnwidth]{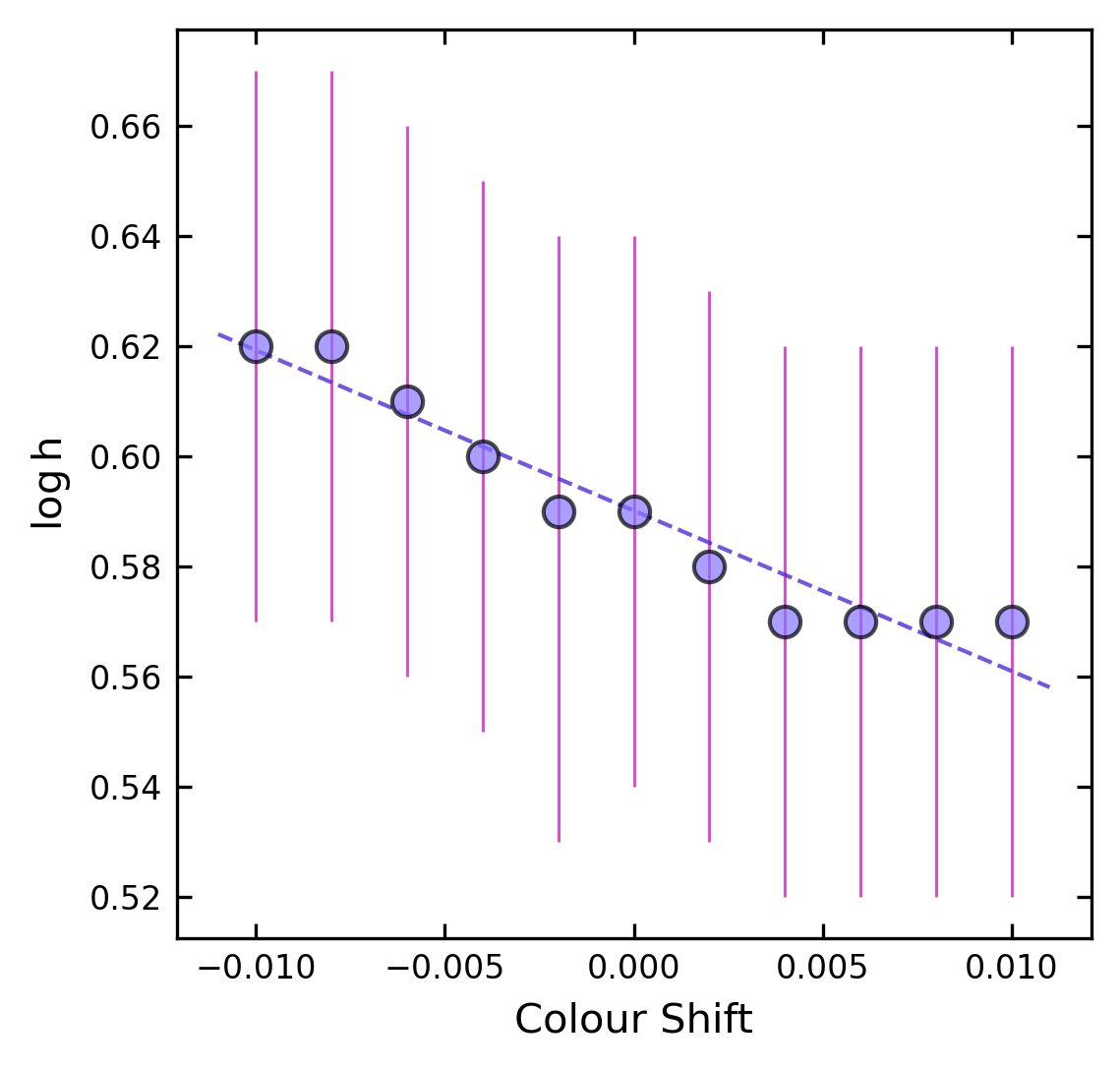}
     \end{subfigure}
        \caption{Top left panel: CMD of Pleiades. The green dashed-line indicates the isochrone used in the analysis. A range of tested
        isochrone shifts in $G-RP$ is indicated with the magenta dashed-lines. Remaining panels show the sensitivity
        of $f_B(q>0.5)$, $FQ_{75}$, and data error scaling factor ($h$) to shifts in $G-RP$.}
        \label{Fb_colour_shift}
\end{figure*}

We evaluated how the accuracy of positioning the isochrones effected our results. To do this, we applied a constant colour shift to our isochrones then applied our probabilistic generative model. We consider a colour shift of $\Delta(G-RP)=\pm0.01$ to be a reasonable upper bound for the uncertainty in fitting our isochrones. The range of this colour shift is displayed for the CMD of Pleiades in Figure \ref{Fb_colour_shift}. We ran our model for the colour shift of $\pm0.01$ in incremental shifts of 0.002 such that we have ten distinct points and an unshifted control case. In Figure \ref{Fb_colour_shift}, for each positive and negative incremental shift in colour applied to the isochrone of Pleiades, we plot each of the binary fraction (for $q\geq0.5$), $FQ_{75}$, and $h$ with 1$\sigma$ uncertainties. A displacement of the isochrone from the cluster main-sequence ridge line is partially compensated by an increase in data error-bar scaling. All of the points lie well within the quoted uncertainties for each panel from the figure and show that our model is extremely robust to the position of the isochrone.

\subsection{Sensitivity to noise}

\begin{figure}
    \centering
    \includegraphics[width=\columnwidth]{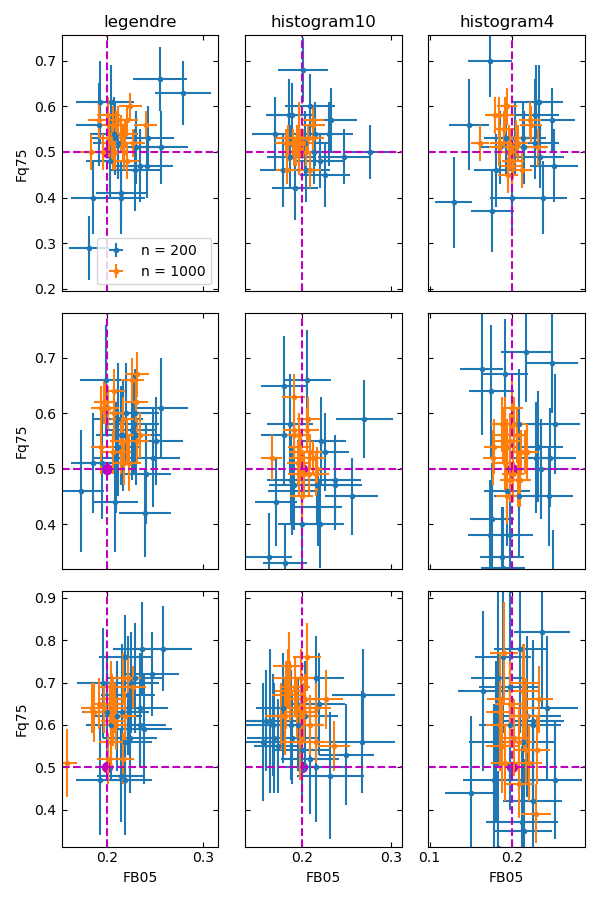}
    \caption{$FQ_{75}$ vs $f_B(q>0.5)$ from simulated CMDs with 200 or 1000 stars and a flat binary mass-ratio distribution. Simulated stars were  given data uncertainties and gaussian scatter the same as Pleiades (for a given $G$). Columns are from analyses with mass-ratio distribution functions represented by Legendre polynomials and histograms with 10 and 4 bins. Rows represent data uncertainties and scatter scaled by $h = (1, 3, 10)$ from top to bottom.}
    \label{fig:sims}
\end{figure}

To determine our sensitivity to noisy data, we have run a number of simulations using a flat
binary mass ratio distribution, and an overall binary fraction of 0.4. The truth value for these simulations
is therefore $f_B(q\geq 0.5) = 0.2$ and $FQ_{75} = 0.5$. We ran simulations for $N = 200$ and $N = 1000$ total stars, initially adopting the mean data uncertainty for each 1-mag bin in $M_G$ as Pleiades. We also ran simulations
with data having 3 and 10 times this noise level. We then modelled the simulated data using the
Legendre basis functions for $P(q)$, and also using histogram basis functions with $N_q = 4$ and 
$N_q = 10$. 20 such simulations were run for each combination of basis functions, noise level, and N.
The results are shown in Figure~\ref{fig:sims}. 

To compare these results with the observed clusters, we adopt the mean uncertainty in $G-RP$ for
$8.0 < M_G < 8.5$ as a metric. The 3 rows in Figure~\ref{fig:sims} have values of this metric of
0.003, 0.009, and 0.03 respectively, while the values for the observed clusters, along with the number of stars considered, are (Col 69, 0.022, 529), ($\alpha $ Per, 0.008, 176), (Pleiades, 0.010, 275), (NGC 6405, 0.025, 392), (Trumpler 10, 0.011, 267),
(UKP 640, 0.027, 276).
 The observed clusters (apart from $\alpha$ Persei) thus have noise characteristics that lie between the second and third rows in Figure~\ref{fig:sims}, with an expected scatter between the orange and blue points. 

From the simulations we can see that there is a bias in the $FQ_{75}$ parameter that develops
in the second row for the Legendre basis functions and the third row for the histogram basis functions. 
The origin of this bias is the non-linear way that $q$ maps to colour on the CMD.  
The transformation from $(M, q)$ to the CMD is unique, but that is not true for the reverse transformation.
For the histogram basis, the overall scatter is lower for $N_q = 10$ than $N_q = 4$. We thus adopt the
$N_q = 10$ histogram basis as being the least sensitive to data noise and expect that the noisiest three clusters in our sample (Col 69, NGC 6405, and UPK 640) may have a measured
$FQ_{75}$ that is too high by approximately 0.15. The other clusters should have little or no bias.
A noise corrected $FQ_\mathrm{75}$ has been plotted for the nosiest clusters in Figure \ref{LH4_10Bins} in the row for Legendre basis functions. This would also greatly effect the trends and could be interpreted as scatter.


\subsection{Comparisons}

There are two points to keep in mind when we make comparisons to results from other papers. Firstly, there is a variation in the mass-ratio range amongst papers when quoting their result for the binary fraction of a star cluster. Secondly, authors will cut off their data at different magnitudes hence there can be some slight variation in the dataset even when comparing work for the same cluster. In Table \ref{q values} we include the binary fraction for a range of mass-ratio values to account for this first issue and the mass range to account for the second issue.


\subsubsection{Collinder 69}

Collinder 69 is the youngest cluster among those examined in this paper. It displays the highest binary fraction of $\mathrm{f_B(q\geq0.5)=0.272^{+0.034}_{-0.036}}$ from our sample. In a comparable study by \cite{pang2023}, a larger binary fraction of $f_B(q\geq0.4) =33.9\pm1.4\%$ was reported before a completeness correction, which subsequently increased to $f_B(q\geq0.4)=48.5\pm2.1\%$, denoted as their upper limit. Our analysis over the same mass range shows a comparable binary fraction of $\mathrm{f_B(q\geq0.4)=0.341^{+0.036}_{-0.038}}$.

\subsubsection{$\alpha$ Persei}

We found a binary fraction of $\alpha$ Persei to be $\mathrm{f_B(q\geq0.5)=0.211^{+0.040}_{-0.039}}$. \cite{Makarov_2006} reported a similarly low binary fraction, noting that approximately 20\% of cluster members were identified as known or suspected spectroscopic, astrometric, or visual binaries or multiple systems. Furthermore, \cite{malofeeva2023} determined a multiplicity fraction of $0.48\pm0.02$ for $q>0.2$, with the combined ratios of triples and quadruples constituting approximately $0.13$ of the total multiplicity. Using Table 1 from their paper we can estimate the binary fraction to be 20-25\% for $q>0.4$, which is consistent with our result of $\mathrm{f_B(q\geq0.4)=0.262^{+0.052}_{-0.050}}$.

\subsubsection{Pleiades}

The determined value of $\mathrm{f_B(q\geq0.5)=0.174^{+0.019}_{-0.022}}$ represents the smallest among the selected young open clusters. Pleiades' binary fraction, as reported in \cite{Donada_2023}, is significantly smaller at $\mathrm{f_B(q>0.6)=0.086^{+0.012}_{-0.011}}$. This is notably less than the values reported by \cite{Niu2020} ($\mathrm{f_B(q<1.0)=0.21\pm0.03}$) and \cite{Pinfield2003} ($\mathrm{f_B(q \geq 0.5)=0.23^{+0.06}_{-0.05}}$), the former of which spans a different mass range. \cite{Jadhav_2021} found the binary fraction to be $\mathrm{f_B(q>0.6)=0.14\pm0.02}$, and over the relevant mass-ratio range, our study determines $\mathrm{f_B(q\geq0.6)=0.127^{+0.017}_{-0.020}}$.

\subsubsection{NGC 6405}

We find the binary fraction of NGC 6405 to be $\mathrm{f_B(q\geq0.5)=0.223^{+0.021}_{-0.023}}$. In \cite{pang2023} the binary fraction 
was determined to be $\mathrm{f_B(q>0.4)=0.158\pm0.018}$. This can also be compared to the average value of $\mathrm{f_B(q>0.6)=0.17\pm0.02}$ from \cite{Jadhav_2021} which aligns with our result at that mass-ratio of $\mathrm{f_B(q\geq0.6)=0.163^{+0.019}_{-0.021}}$.

\subsubsection{Trumpler 10}

We found a binary fraction of $\mathrm{f_B(q\geq0.5)=0.186^{+0.020}_{-0.022}}$ for this cluster although the value of $\mathrm{f_B(q\geq0.6)=0.124^{+0.018}_{-0.020}}$ can be considered for comparisons. This is smaller than $\mathrm{f_{B}\left(q>0.6^{+0.05}_{-0.15}\right)=0.159^{+0.019}_{-0.018}}$ of \cite{Donada2022}, but similar to $\mathrm{f_B(q>0.6)=0.12\pm0.02}$ from \cite{Jadhav_2021}.

\subsubsection{UPK 640}

For UPK 640 we found $\mathrm{f_B(q\geq0.5)=0.214^{+0.039}_{-0.040}}$, although for comparisons we shall consider $\mathrm{f_B(q\geq0.4)=0.274^{+0.048}_{-0.050}}$ which is the third highest from our study. This differs from \cite{pang2023}, who reported $\mathrm{f_{B}(q>0.4)=0.104\pm0.005}$ and $\mathrm{f_{B}(q>0.4)=0.166\pm0.010}$ before and after applying a completeness correction (note that the cluster is referred to as LP 2442 in their work.) In the UPK 640 region, \cite{Pang_2022} identified five hierarchical groups, and \cite{pang2023} determined their binary fraction after completeness correction to range from $\mathrm{f_{B}(q>0.4)=0.020\pm0.002}$ to $\mathrm{f_{B}(q>0.4)=0.44\pm0.04}$.

\section{Discussion}

The major findings from our analysis are (i) that the fraction of binary stars (at least those with $q > 0.5$) declines with cluster age,
 and (ii) that the proportion  of binaries with high mass-ratios ($q > 0.75$) increases with the overall fraction of binaries. The first of these trends has been previously reported, for instance by \citet{Sollima_2010} 
 and \citet{Donada_2023}, and is consistent with the models of \citet{Sollima2008}, which predict an overall decline in the frequency of binaries over the first Gyr of a cluster's life.

The second trend is more difficult to explain. It indicates that high-$q$ binaries are lost preferentially with age, compared to all binaries.
If the loss of  high-$q$ binaries with age  were due to preferential ejection of these systems as a cluster evolves,
we would expect to find more of these binaries outside the cluster cores. However, this would contradict
\citet{Albrow23}, which found that binaries retained in the cores of Hyades and Praesepe had a higher mass-ratio than those being ejected from the clusters.

An alternative explanation may lie in the observation \citep{Duquennoy1991} that binaries in tight orbits tend to have mass-ratios close to unity. 
\citet{Duquennoy1991} also note that the radial velocity measurements of binaries in the Hyades by \citet{Griffin1985} suggest an excess of short-period binaries
in open clusters compared with the field. 
Indeed, the primordial period distribution of binaries in clusters may be closer to
uniform in log-period rather than the log-normal distribution of field binaries
\citep{Connelley2008, Reipurth2007}. 
Applying Bayesian reasoning to these two observations, we can infer that cluster binaries with high mass-ratios are more likely to be tightly bound.
Coupled with our observation that $FQ_{75}$ declines with age, this suggests an excess population of primordial close binaries that declines with age. 

The orbits of close binaries can decay via tidal friction \citep{Zahn1977} and
the stars can eventually merge.
However, the theoretical models
predict a merger time scale of 1 Gyr \citep{Stepien1995} or much longer \citep{Andronov2006} for low mass binaries, assuming initial orbital period of 2 days or longer. 

This points to internal dynamical processing as being a significant factor for the observed trends. We note that the models of \citet{Sollima2008} found that collisional disruption of binaries during encounters with a third star (ionization) is the dominant mechanism that controls the fraction of binary stars in clusters.
For this to explain our second trend would imply that collisional disruption
is more likely to affect high-$q$ binary pairs than for those with lower $q$.
This seems unlikely, as we expect collisional disruption to act preferentially
for wide binaries, i.e. those less likely to have an equal mass-ratio.

Consider instead the interaction of a low-mass binary with $q \approx 1$
with a third (single) star. 
In three-body interactions, if the binary system is not dissociated, the most likely result is that
the lowest-mass star of the three is ejected and the remaining two stars
persist as a tight binary system \citep{Heggie1975, Hills1975}.

If the third star is of lower mass that the members of the 
binary, then the original binary system survives, but will generally have
lost angular momentum to the third star, thus hastening its orbital decay and eventual 
coalescence. This will preferentially cause the loss of binaries that are already in close orbits. This channel
thus causes a decline in the {\em number} of close high-$q$ binaries,

If the third star has a higher mass than the stars in the binary,
then one of original binary star members is likely to be ejected
and the new binary system will have a lower mass ratio.
This channel will thus lower the {\em fraction} of binaries with high $q$.

We suggest then that while collisional disruption is likely the cause of the overall reduction in the binary fraction with age, it is non-ionizing interactions
that are responsible for the reduction with cluster age in the fraction of binaries that have high mass-ratios.

\section{Summary}

We have used an adapted version of the probabilistic generative model for colour magnitude diagrams from \cite{Albrow22Ulusele} to measure the mass-ratio distribution and binary fraction of six different young star clusters from Gaia DR3 photometry. 
Collinder 69 exhibits a sharp increase in the mass-ratio distribution at $q\gtrsim0.9$, which is similar to the field-star distribution found by \citet{Fisher2005}. The 5 remaining clusters show a flatter distribution consistent with \citet{Raghav2010}.
Of the binaries in these clusters with $q > 0.5$, $\sim$40-80\% also have $q > 0.75$
The binary frequency was determined for all of these clusters and the results imply that the binary fraction is greater for younger clusters. 
Additionally, the ratio of high-mass-ratio binaries increases with the binary frequency of a cluster, but declines with age. This may be due to an excess primordial population of 
short-period binaries with mass ratios close to one that evolve to lower mass ratios through
three-body interactions.

\section*{Acknowledgements}

We thank the anonymous referee for comments that helped us improve the manuscript.


\section*{Data and Code Availability}

Gaia Data Release 3 (DR3) data, and the Gaia Catalogue of Nearby Stars are publicly available via the Gaia archive,  https://gea.esac.esa.int/archive/ ,
and the Centre de Donn\'ees astronomiques de Strasbourg(CDS) catalogue service, https://vizier.cds.unistra.fr/viz-bin/VizieR.

The CMDFITTER code used for this analysis is written in PYTHON and CUDA (via the pyCUDA python library). 
CUDA is an extension to C/C++ that uses an NVIDIA graphical processing unit to perform
parallel calculations. The code 
is available at \url{https://github.com/MichaelDAlbrow/CMDFitter}.



\bibliographystyle{mnras}
\bibliography{biblio} 

\begin{thebibliography}{}
\makeatletter
\relax
\def\mn@urlcharsother{\let\do\@makeother \do\$\do\&\do\#\do\^\do\_\do\%\do\~}
\def\mn@doi{\begingroup\mn@urlcharsother \@ifnextchar [ {\mn@doi@} {\mn@doi@[]}}
\def\mn@doi@[#1]#2{\def\@tempa{#1}\ifx\@tempa\@empty \href {http://dx.doi.org/#2} {doi:#2}\else \href {http://dx.doi.org/#2} {#1}\fi \endgroup}
\def\mn@eprint#1#2{\mn@eprint@#1:#2::\@nil}
\def\mn@eprint@arXiv#1{\href {http://arxiv.org/abs/#1} {{\tt arXiv:#1}}}
\def\mn@eprint@dblp#1{\href {http://dblp.uni-trier.de/rec/bibtex/#1.xml} {dblp:#1}}
\def\mn@eprint@#1:#2:#3:#4\@nil{\def\@tempa {#1}\def\@tempb {#2}\def\@tempc {#3}\ifx \@tempc \@empty \let \@tempc \@tempb \let \@tempb \@tempa \fi \ifx \@tempb \@empty \def\@tempb {arXiv}\fi \@ifundefined {mn@eprint@\@tempb}{\@tempb:\@tempc}{\expandafter \expandafter \csname mn@eprint@\@tempb\endcsname \expandafter{\@tempc}}}

\bibitem[\protect\citeauthoryear{{Abdurro'uf}, {Accetta}, {Aerts}  et~al.}{{Abdurro'uf} et~al.}{2022}]{Abdurrouf2022}
{Abdurro'uf} {Accetta} K.,  {Aerts} C.,   et~al., 2022, \mn@doi [\apjs] {10.3847/1538-4365/ac4414}, \href {https://ui.adsabs.harvard.edu/abs/2022ApJS..259...35A} {259, 35}

\bibitem[\protect\citeauthoryear{Ahumada \& Lapasset}{Ahumada \& Lapasset}{1995}]{ahumada1995}
Ahumada J.,  Lapasset E.,  1995, \aaps, \href {https://ui.adsabs.harvard.edu/abs/1995A&AS..109..375A} {109, 375}

\bibitem[\protect\citeauthoryear{{Albrow}}{{Albrow}}{2024}]{Albrow23}
{Albrow} M.~D.,  2024, \mn@doi [\mnras] {10.1093/mnras/stae425}, \href {https://ui.adsabs.harvard.edu/abs/2024MNRAS.528.6211A} {528, 6211}

\bibitem[\protect\citeauthoryear{Albrow \& Ulusele}{Albrow \& Ulusele}{2022}]{Albrow22Ulusele}
Albrow M.~D.,  Ulusele I.~H.,  2022, \mn@doi [\mnras] {10.1093/mnras/stac1731}, 515, 730

\bibitem[\protect\citeauthoryear{Albrow, Gilliland, Brown, Edmonds, Guhathakurta  \& Sarajedini}{Albrow et~al.}{2001}]{albrow2001}
Albrow M.~D.,  Gilliland R.~L.,  Brown T.~M.,  Edmonds P.~D.,  Guhathakurta P.,   Sarajedini A.,  2001, \mn@doi [\apj] {10.1086/322353}, 559, 1060

\bibitem[\protect\citeauthoryear{{Andronov}, {Pinsonneault}  \& {Terndrup}}{{Andronov} et~al.}{2006}]{Andronov2006}
{Andronov} N.,  {Pinsonneault} M.~H.,   {Terndrup} D.~M.,  2006, \mn@doi [\apj] {10.1086/505127}, \href {https://ui.adsabs.harvard.edu/abs/2006ApJ...646.1160A} {646, 1160}

\bibitem[\protect\citeauthoryear{Basri \& Martín}{Basri \& Martín}{1999}]{Basri_1999}
Basri G.,  Martín E.~L.,  1999, \mn@doi [\apj] {10.1086/306564}, 510, 266

\bibitem[\protect\citeauthoryear{Bayo, Rodrigo, y Navascu{\'{e} }s, Solano, Guti{\'{e}}rrez, Morales-Calder{\'{o}}n  \& Allard}{Bayo et~al.}{2008}]{Bayo_2008}
Bayo A.,  Rodrigo C.,  y Navascu{\'{e} }s D.~B.,  Solano E.,  Guti{\'{e}}rrez R.,  Morales-Calder{\'{o}}n M.,   Allard F.,  2008, \mn@doi [\aap] {10.1051/0004-6361:200810395}, 492, 277

\bibitem[\protect\citeauthoryear{Bhat, Lanzoni, Ferraro  \& Vesperini}{Bhat et~al.}{2023}]{Bhat_2023}
Bhat B.,  Lanzoni B.,  Ferraro F.~R.,   Vesperini E.,  2023, \mn@doi [\apj] {10.3847/1538-4357/acb434}, 945, 164

\bibitem[\protect\citeauthoryear{{Brandner}, {Calissendorff}  \& {Kopytova}}{{Brandner} et~al.}{2023}]{Brandner2023}
{Brandner} W.,  {Calissendorff} P.,   {Kopytova} T.,  2023, \mn@doi [\mnras] {10.1093/mnras/stac2247}, \href {https://ui.adsabs.harvard.edu/abs/2023MNRAS.518..662B} {518, 662}

\bibitem[\protect\citeauthoryear{Cantat-Gaudin et~al.}{Cantat-Gaudin et~al.}{2020}]{Cantat}
Cantat-Gaudin T.,  et~al., 2020, \mn@doi [\aap] {10.1051/0004-6361/202038192}, \href {https://ui.adsabs.harvard.edu/abs/2020A&A...640A...1C} {640, A1}

\bibitem[\protect\citeauthoryear{Chabrier}{Chabrier}{2003}]{Chabrier_2003}
Chabrier G.,  2003, \mn@doi [\pasp] {10.1086/376392}, 115, 763

\bibitem[\protect\citeauthoryear{{Choi} et~al.}{{Choi} et~al.}{2016}]{MIST2}
{Choi} J.,  et~al., 2016, \mn@doi [\apj] {10.3847/0004-637X/823/2/102}, \href {https://ui.adsabs.harvard.edu/abs/2016ApJ...823..102C} {823, 102}

\bibitem[\protect\citeauthoryear{{Connelley}, {Reipurth}  \& {Tokunaga}}{{Connelley} et~al.}{2008}]{Connelley2008}
{Connelley} M.~S.,  {Reipurth} B.,   {Tokunaga} A.~T.,  2008, \mn@doi [\aj] {10.1088/0004-6256/135/6/2526}, \href {https://ui.adsabs.harvard.edu/abs/2008AJ....135.2526C} {135, 2526}

\bibitem[\protect\citeauthoryear{{Cordoni} et~al.,}{{Cordoni} et~al.}{2023}]{Cordoni2023}
{Cordoni} G.,  et~al., 2023, \mn@doi [\aap] {10.1051/0004-6361/202245457}, \href {https://ui.adsabs.harvard.edu/abs/2023A&A...672A..29C} {672, A29}

\bibitem[\protect\citeauthoryear{{Dahm}}{{Dahm}}{2015}]{Dahm2015}
{Dahm} S.~E.,  2015, \mn@doi [\apj] {10.1088/0004-637X/813/2/108}, \href {https://ui.adsabs.harvard.edu/abs/2015ApJ...813..108D} {813, 108}

\bibitem[\protect\citeauthoryear{Dolan \& Mathieu}{Dolan \& Mathieu}{1999}]{Dolan_1999}
Dolan C.~J.,  Mathieu R.~D.,  1999, \mn@doi [\apj] {10.1086/301075}, 118, 2409

\bibitem[\protect\citeauthoryear{Donada}{Donada}{2022}]{Donada2022}
Donada J.,  2022, Master's thesis, Universitat de Barcelona

\bibitem[\protect\citeauthoryear{Donada et~al.,}{Donada et~al.}{2023}]{Donada_2023}
Donada J.,  et~al., 2023, \mn@doi [\aap] {10.1051/0004-6361/202245219}, 675, A89

\bibitem[\protect\citeauthoryear{{Dotter}}{{Dotter}}{2016}]{MIST1}
{Dotter} A.,  2016, \mn@doi [\apjs] {10.3847/0067-0049/222/1/8}, \href {https://ui.adsabs.harvard.edu/abs/2016ApJS..222....8D} {222, 8}

\bibitem[\protect\citeauthoryear{{Duquennoy} \& {Mayor}}{{Duquennoy} \& {Mayor}}{1991a}]{Duq91}
{Duquennoy} A.,  {Mayor} M.,  1991a, \aap, \href {https://ui.adsabs.harvard.edu/abs/1991A&A...248..485D} {248, 485}

\bibitem[\protect\citeauthoryear{{Duquennoy} \& {Mayor}}{{Duquennoy} \& {Mayor}}{1991b}]{Duquennoy1991}
{Duquennoy} A.,  {Mayor} M.,  1991b, \aap, \href {https://ui.adsabs.harvard.edu/abs/1991A&A...248..485D} {248, 485}

\bibitem[\protect\citeauthoryear{{Elsanhoury}, Postnikova, Chupina  et~al.}{{Elsanhoury} et~al.}{2018}]{Elsanhoury15}
{Elsanhoury} W.~H.,  Postnikova E.~S.,  Chupina N.~V.,   et~al., 2018, \mn@doi [\apss] {101007/s10509-018-3268-3}, \href {http://dx.doi.org/10.1007/s10509-018-3268-3} {363, 58}

\bibitem[\protect\citeauthoryear{{Fisher}, {Schr{\"o}der}  \& {Smith}}{{Fisher} et~al.}{2005}]{Fisher2005}
{Fisher} J.,  {Schr{\"o}der} K.-P.,   {Smith} R.~C.,  2005, \mn@doi [\mnras] {10.1111/j.1365-2966.2005.09193.x}, \href {https://ui.adsabs.harvard.edu/abs/2005MNRAS.361..495F} {361, 495}

\bibitem[\protect\citeauthoryear{Fürnkranz, Rix, Coronado  \& Seeburger}{Fürnkranz et~al.}{2023}]{fürnkranz2023}
Fürnkranz V.,  Rix H.-W.,  Coronado J.,   Seeburger R.,  2023, \mn@doi [\apj] {10.48550/arXiv.2311.10133}

\bibitem[\protect\citeauthoryear{{Griffin}}{{Griffin}}{1985}]{Griffin1985}
{Griffin} R.~F.,  1985, in {Eggleton} P.~P.,  {Pringle} J.~E.,  eds,  NATO Advanced Study Institute (ASI) Series C Vol. 150, Interacting Binaries. p.~1, \mn@doi{10.1007/978-94-009-5337-6_1}

\bibitem[\protect\citeauthoryear{Healy, McCullough, Schlaufman  \& Kovacs}{Healy et~al.}{2023}]{Healy_2023}
Healy B.~F.,  McCullough P.~R.,  Schlaufman K.~C.,   Kovacs G.,  2023, \mn@doi [\apj] {10.3847/1538-4357/acad7b}, 944, 39

\bibitem[\protect\citeauthoryear{{Heggie}}{{Heggie}}{1975}]{Heggie1975}
{Heggie} D.~C.,  1975, \mn@doi [\mnras] {10.1093/mnras/173.3.729}, \href {https://ui.adsabs.harvard.edu/abs/1975MNRAS.173..729H} {173, 729}

\bibitem[\protect\citeauthoryear{{Higson}, {Handley}, {Hobson}  \& {Lasenby}}{{Higson} et~al.}{2019}]{DynamicSampling}
{Higson} E.,  {Handley} W.,  {Hobson} M.,   {Lasenby} A.,  2019, \mn@doi [Statistics and Computing] {10.1007/s11222-018-9844-0}, \href {https://ui.adsabs.harvard.edu/abs/2019S&C....29..891H} {29, 891}

\bibitem[\protect\citeauthoryear{{Hills}}{{Hills}}{1975}]{Hills1975}
{Hills} J.~G.,  1975, \mn@doi [\aj] {10.1086/111815}, \href {https://ui.adsabs.harvard.edu/abs/1975AJ.....80..809H} {80, 809}

\bibitem[\protect\citeauthoryear{Hurley, Pols, Aarseth  \& Tout}{Hurley et~al.}{2005}]{Hur2005}
Hurley J.,  Pols O.,  Aarseth S.,   Tout C.,  2005, \mn@doi [\mnras] {10.1111/j.1365-2966.2005.09448.x}, 363

\bibitem[\protect\citeauthoryear{Jadhav, Roy, Joshi  \& Subramaniam}{Jadhav et~al.}{2021}]{Jadhav_2021}
Jadhav V.~V.,  Roy K.,  Joshi N.,   Subramaniam A.,  2021, \mn@doi [\apj] {10.3847/1538-3881/ac2571}, 162, 264

\bibitem[\protect\citeauthoryear{Kim, Walter  \& Wolk}{Kim et~al.}{2005}]{kim2005}
Kim J.~S.,  Walter F.~M.,   Wolk S.~J.,  2005, \mn@doi [\apj] {10.1086/428002}, 129, 1564

\bibitem[\protect\citeauthoryear{Koen}{Koen}{2001}]{Koentime}
Koen M.~C.,  2001, Ph{D} thesis, University of Johannesburg

\bibitem[\protect\citeauthoryear{{Kroupa}, {Tout}  \& {Gilmore}}{{Kroupa} et~al.}{1990}]{Kroupa90}
{Kroupa} P.,  {Tout} C.~A.,   {Gilmore} G.,  1990, \mnras, \href {https://ui.adsabs.harvard.edu/abs/1990MNRAS.244...76K} {244, 76}

\bibitem[\protect\citeauthoryear{{Li}, {Shao}, {Li}, {Yu}, {Zhong}  \& {Chen}}{{Li} et~al.}{2020}]{Li2020}
{Li} L.,  {Shao} Z.,  {Li} Z.-Z.,  {Yu} J.,  {Zhong} J.,   {Chen} L.,  2020, \mn@doi [\apj] {10.3847/1538-4357/abaef3}, \href {https://ui.adsabs.harvard.edu/abs/2020ApJ...901...49L} {901, 49}

\bibitem[\protect\citeauthoryear{Li et~al.,}{Li et~al.}{2022}]{Li_2022}
Li J.,  et~al., 2022, \mn@doi [\apj] {10.3847/1538-4357/ac731d}, 933, 119

\bibitem[\protect\citeauthoryear{{Lindoff}}{{Lindoff}}{1968}]{Lindoff68}
{Lindoff} U.,  1968, Ark. Astr, \href {https://ui.adsabs.harvard.edu/abs/1968ArA.....5....1L} {5, 1}

\bibitem[\protect\citeauthoryear{Makarov}{Makarov}{2006}]{Makarov_2006}
Makarov V.~V.,  2006, \mn@doi [\apj] {10.1086/503900}, 131, 2967

\bibitem[\protect\citeauthoryear{Malofeeva, Mikhnevich, Carraro  \& Seleznev}{Malofeeva et~al.}{2023}]{malofeeva2023}
Malofeeva A.~A.,  Mikhnevich V.~O.,  Carraro G.,   Seleznev A.~F.,  2023, \mn@doi [\apj] {10.3847/1538-3881/aca666}, 165, 45

\bibitem[\protect\citeauthoryear{{Mermilliod}}{{Mermilliod}}{1981}]{Mermilliod81}
{Mermilliod} J.~C.,  1981, \aap, \href {https://ui.adsabs.harvard.edu/abs/1981A&A....97..235M} {97, 235}

\bibitem[\protect\citeauthoryear{Mermilliod, Queloz  \& Mayor}{Mermilliod et~al.}{2008}]{mermilliod2008}
Mermilliod J.-C.,  Queloz D.,   Mayor M.,  2008, \mn@doi [\aap] {https://doi.org/10.1051/0004-6361:200810016}, 488, 409

\bibitem[\protect\citeauthoryear{Milone et~al.,}{Milone et~al.}{2012}]{Milone_2012}
Milone A.~P.,  et~al., 2012, \mn@doi [\aap] {10.1051/0004-6361/201016384}, 540, A16

\bibitem[\protect\citeauthoryear{Moe \& Stefano}{Moe \& Stefano}{2017}]{Moe_2017}
Moe M.,  Stefano R.~D.,  2017, \mn@doi [\apjs] {10.3847/1538-4365/aa6fb6}, 230, 15

\bibitem[\protect\citeauthoryear{Nikiforova, Kulesh, Seleznev  \& Carraro}{Nikiforova et~al.}{2020}]{Nikiforova_2020}
Nikiforova V.~V.,  Kulesh M.~V.,  Seleznev A.~F.,   Carraro G.,  2020, \mn@doi [\apj] {10.3847/1538-3881/aba753}, 160, 142

\bibitem[\protect\citeauthoryear{{Niu}, {Wang}  \& {Fu}}{{Niu} et~al.}{2020}]{Niu2020}
{Niu} H.,  {Wang} J.,   {Fu} J.,  2020, \mn@doi [\apj] {10.3847/1538-4357/abb8d6}, \href {https://ui.adsabs.harvard.edu/abs/2020ApJ...903...93N} {903, 93}

\bibitem[\protect\citeauthoryear{{Pamos Ortega, D.}, {Mirouh, G. M.}, {Garc\'{\i}a Hern\'andez, A.}, {Su\'arez Yanes, J. C.}  \& {Barcel\'o Forteza, S.}}{{Pamos Ortega, D.} et~al.}{2023}]{Ortega23}
{Pamos Ortega, D.} {Mirouh, G. M.} {Garc\'{\i}a Hern\'andez, A.} {Su\'arez Yanes, J. C.}  {Barcel\'o Forteza, S.} 2023, \mn@doi [\aap] {10.1051/0004-6361/202346323}, 675, A167

\bibitem[\protect\citeauthoryear{Pang et~al.,}{Pang et~al.}{2022}]{Pang_2022}
Pang X.,  et~al., 2022, \mn@doi [\apj] {10.3847/1538-4357/ac674e}, 931, 156

\bibitem[\protect\citeauthoryear{Pang et~al.,}{Pang et~al.}{2023}]{pang2023}
Pang X.,  et~al., 2023, \mn@doi [\apj] {10.3847/1538-3881/ace76c}, 166, 110

\bibitem[\protect\citeauthoryear{Paunzen, Netopil, Iliev, Maitzen, Claret  \& Pintado}{Paunzen et~al.}{2006}]{paunzen2006}
Paunzen E.,  Netopil M.,  Iliev I.~K.,  Maitzen H.,  Claret A.,   Pintado O.,  2006, \mn@doi [\aap] {10.1051/0004-6361:20054628}, 454, 171

\bibitem[\protect\citeauthoryear{{Paxton} et~al.}{{Paxton} et~al.}{2018}]{MIST3}
{Paxton} B.,  et~al., 2018, \mn@doi [\apjs] {10.3847/1538-4365/aaa5a8}, \href {https://ui.adsabs.harvard.edu/abs/2018ApJS..234...34P} {234, 34}

\bibitem[\protect\citeauthoryear{Pinfield, Dobbie, Jameson, Steele, Jones  \& Katsiyannis}{Pinfield et~al.}{2003}]{Pinfield2003}
Pinfield D.~J.,  Dobbie P.~D.,  Jameson R.~F.,  Steele I.~A.,  Jones H. R.~A.,   Katsiyannis A.~C.,  2003, \mn@doi [\mnras] {10.1046/j.1365-8711.2003.06630.x}, 342, 1241

\bibitem[\protect\citeauthoryear{{Raghavan} et~al.,}{{Raghavan} et~al.}{2010}]{Raghav2010}
{Raghavan} D.,  et~al., 2010, \mn@doi [\apjs] {10.1088/0067-0049/190/1/1}, \href {https://ui.adsabs.harvard.edu/abs/2010ApJS..190....1R} {190, 1}

\bibitem[\protect\citeauthoryear{{Reipurth}, {Guimar{\~a}es}, {Connelley}  \& {Bally}}{{Reipurth} et~al.}{2007}]{Reipurth2007}
{Reipurth} B.,  {Guimar{\~a}es} M.~M.,  {Connelley} M.~S.,   {Bally} J.,  2007, \mn@doi [\aj] {10.1086/523596}, \href {https://ui.adsabs.harvard.edu/abs/2007AJ....134.2272R} {134, 2272}

\bibitem[\protect\citeauthoryear{{Salpeter}}{{Salpeter}}{1955}]{Salpeter55}
{Salpeter} E.~E.,  1955, \mn@doi [\apj] {10.1086/145971}, \href {https://ui.adsabs.harvard.edu/abs/1955ApJ...121..161S} {121, 161}

\bibitem[\protect\citeauthoryear{Sim, Lee, Ann  \& Kim}{Sim et~al.}{2019}]{Sim2019}
Sim G.,  Lee S.~H.,  Ann H.~B.,   Kim S.,  2019, \mn@doi [J. Korean Astron. Soc.] {10.5303/JKAS.2019.52.5.145}, 52, 145

\bibitem[\protect\citeauthoryear{Soderblom, Laskar, Valenti, Stauffer  \& Rebull}{Soderblom et~al.}{2009}]{Soderblom_2009}
Soderblom D.~R.,  Laskar T.,  Valenti J.~A.,  Stauffer J.~R.,   Rebull L.~M.,  2009, \mn@doi [\apj] {10.1088/0004-6256/138/5/1292}, 138, 1292

\bibitem[\protect\citeauthoryear{{Sollima}}{{Sollima}}{2008}]{Sollima2008}
{Sollima} A.,  2008, \mn@doi [\mnras] {10.1111/j.1365-2966.2008.13387.x}, \href {https://ui.adsabs.harvard.edu/abs/2008MNRAS.388..307S} {388, 307}

\bibitem[\protect\citeauthoryear{Sollima, Carballo-Bello, Beccari, Ferraro, Pecci  \& Lanzoni}{Sollima et~al.}{2010}]{Sollima_2010}
Sollima A.,  Carballo-Bello J.~A.,  Beccari G.,  Ferraro F.~R.,  Pecci F.~F.,   Lanzoni B.,  2010, \mn@doi [\mnras] {10.1111/j.1365-2966.2009.15676.x}, 401, 577

\bibitem[\protect\citeauthoryear{{Speagle}}{{Speagle}}{2020}]{DYNESTY}
{Speagle} J.~S.,  2020, \mn@doi [\mnras] {10.1093/mnras/staa278}, \href {https://ui.adsabs.harvard.edu/abs/2020MNRAS.493.3132S} {493, 3132}

\bibitem[\protect\citeauthoryear{{Stepien}}{{Stepien}}{1995}]{Stepien1995}
{Stepien} K.,  1995, \mn@doi [\mnras] {10.1093/mnras/274.4.1019}, \href {https://ui.adsabs.harvard.edu/abs/1995MNRAS.274.1019S} {274, 1019}

\bibitem[\protect\citeauthoryear{{Tout}}{{Tout}}{1991}]{Tout91}
{Tout} C.~A.,  1991, \mn@doi [\mnras] {10.1093/mnras/250.4.701}, \href {https://ui.adsabs.harvard.edu/abs/1991MNRAS.250..701T} {250, 701}

\bibitem[\protect\citeauthoryear{{Vleeming}}{{Vleeming}}{1974}]{Vleeming74}
{Vleeming} G.,  1974, \aaps, \href {https://ui.adsabs.harvard.edu/abs/1974A&AS...16..331V} {16, 331}

\bibitem[\protect\citeauthoryear{{Yalyalieva}, {Carraro}, {Glushkova}, {Munari}  \& {Ochner}}{{Yalyalieva} et~al.}{2022}]{Yalyalieva2022}
{Yalyalieva} L.,  {Carraro} G.,  {Glushkova} E.,  {Munari} U.,   {Ochner} P.,  2022, \mn@doi [\mnras] {10.1093/mnras/stac1199}, \href {https://ui.adsabs.harvard.edu/abs/2022MNRAS.513.5299Y} {513, 5299}

\bibitem[\protect\citeauthoryear{{Yalyalieva}, {Chemel}, {Carraro}  \& {Glushkova}}{{Yalyalieva} et~al.}{2024}]{Yalyalieva2024}
{Yalyalieva} L.,  {Chemel} A.,  {Carraro} G.,   {Glushkova} E.,  2024, \mn@doi [\aj] {10.3847/1538-3881/ad1ff0}, \href {https://ui.adsabs.harvard.edu/abs/2024AJ....167..100Y} {167, 100}

\bibitem[\protect\citeauthoryear{{Zahn}}{{Zahn}}{1977}]{Zahn1977}
{Zahn} J.~P.,  1977, \aap, \href {https://ui.adsabs.harvard.edu/abs/1977A&A....57..383Z} {57, 383}

\makeatother
\end{thebibliography}



\appendix

\section{Realisations}
\begin{figure*}
    \centering
    \begin{subfigure}[b]{0.95\textwidth}
        \centering
        \includegraphics[width=\columnwidth]{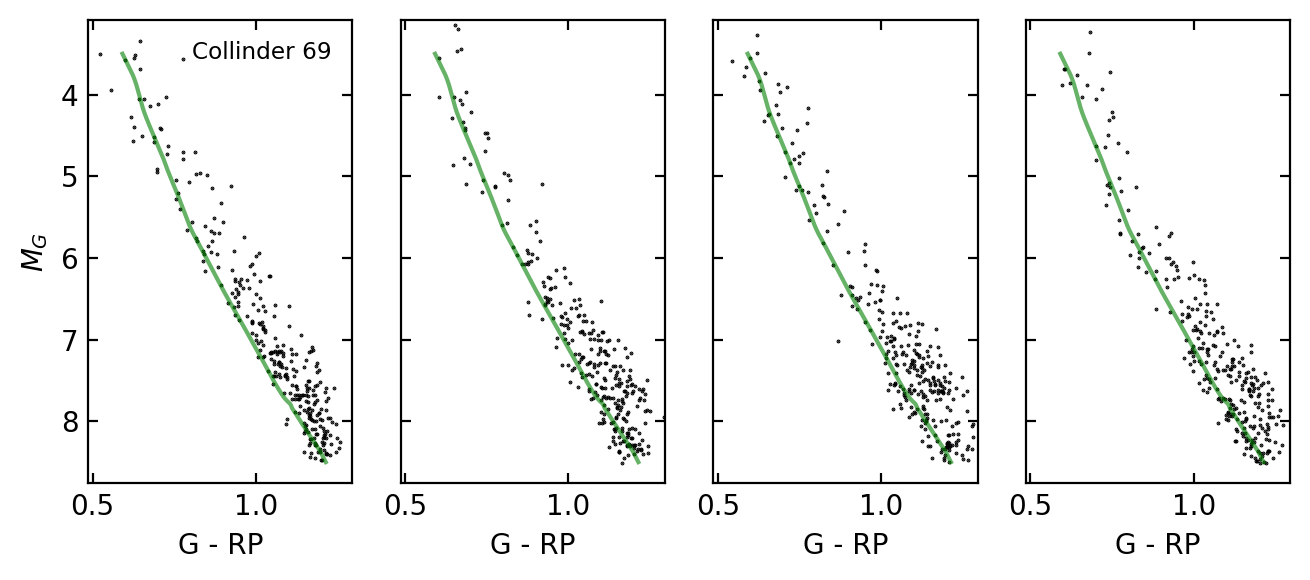}
    \end{subfigure}
    \hfill
    \begin{subfigure}[b]{0.95\textwidth}
        \centering
        \includegraphics[width=\columnwidth]{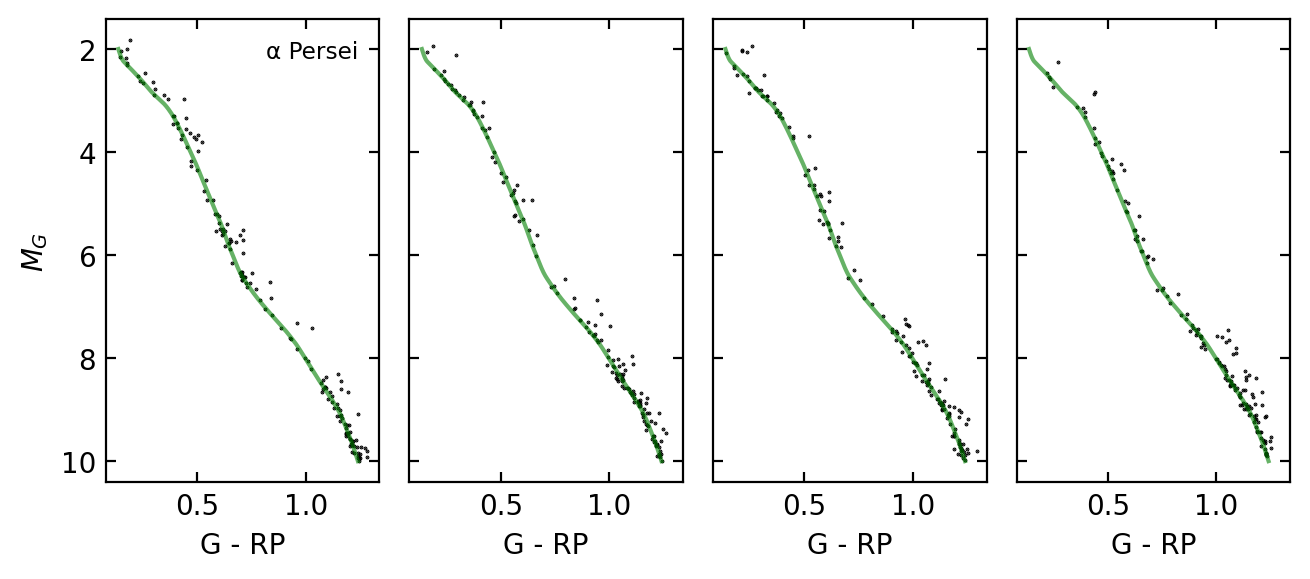}
    \end{subfigure}
    \hfill
    \begin{subfigure}[b]{0.95\textwidth}
        \centering
        \includegraphics[width=\columnwidth]{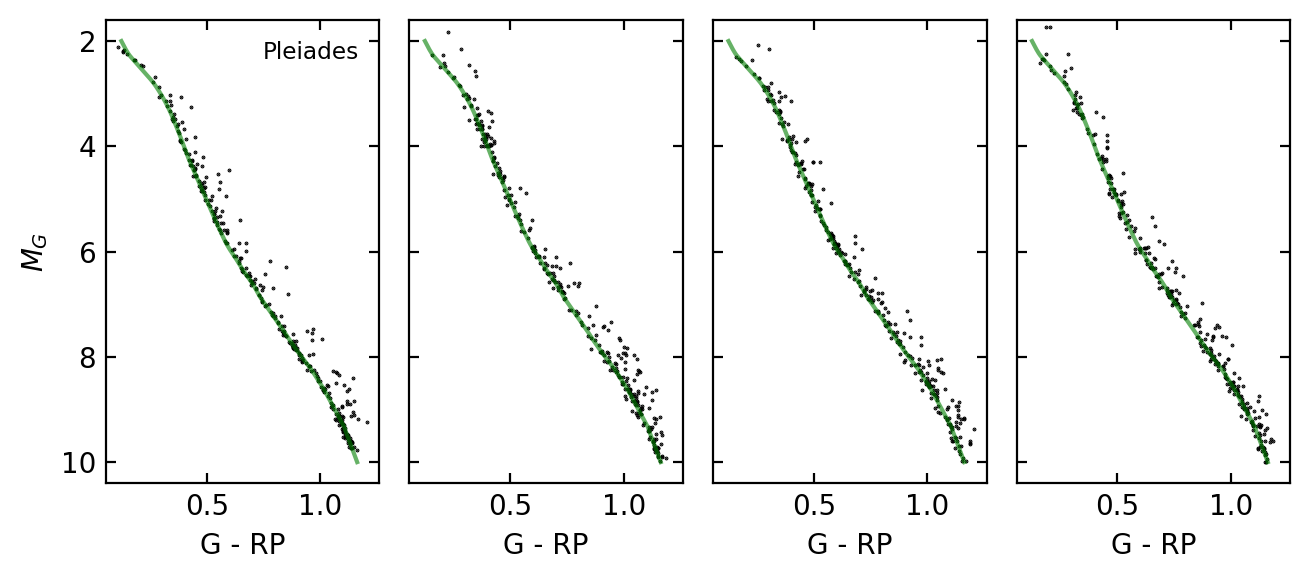}
    \end{subfigure}
    \caption{Observed CMD for each cluster, accompanied by 3 random realisations from the posterior maximum model for the Legendre 
    mass-ratio distribution representation.}
    \label{Realisation}
\end{figure*}

\begin{figure*}
    \centering
    \begin{subfigure}[b]{0.95\textwidth}
        \centering
        \includegraphics[width=\columnwidth]{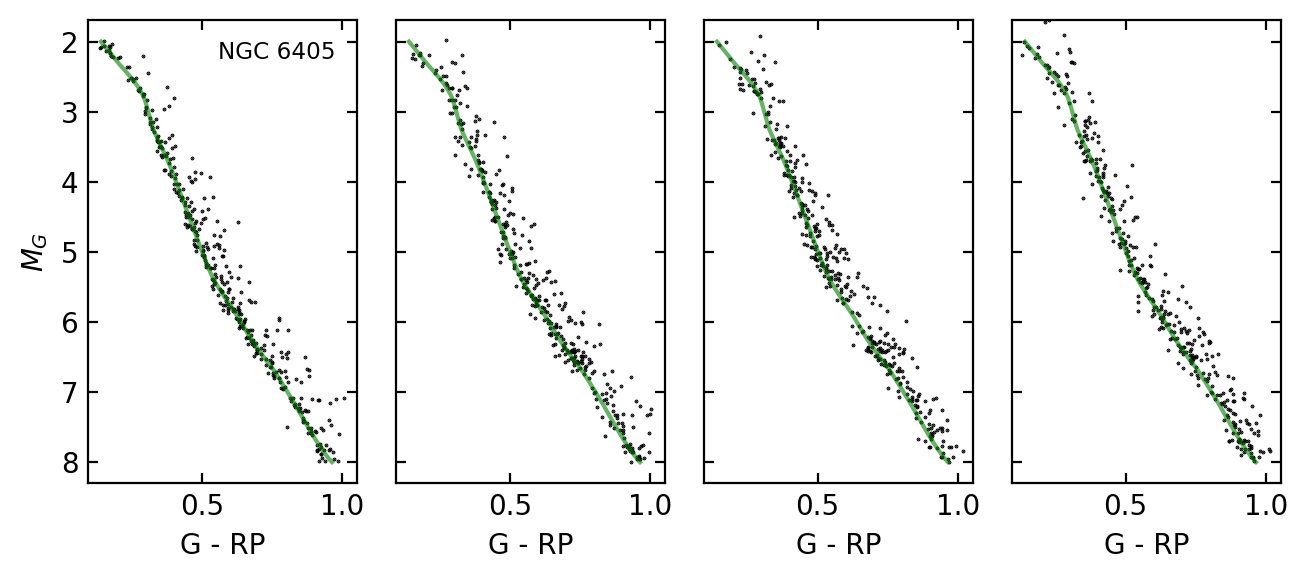}
    \end{subfigure}
    \hfill
    \begin{subfigure}[b]{0.95\textwidth}
        \centering
        \includegraphics[width=\columnwidth]{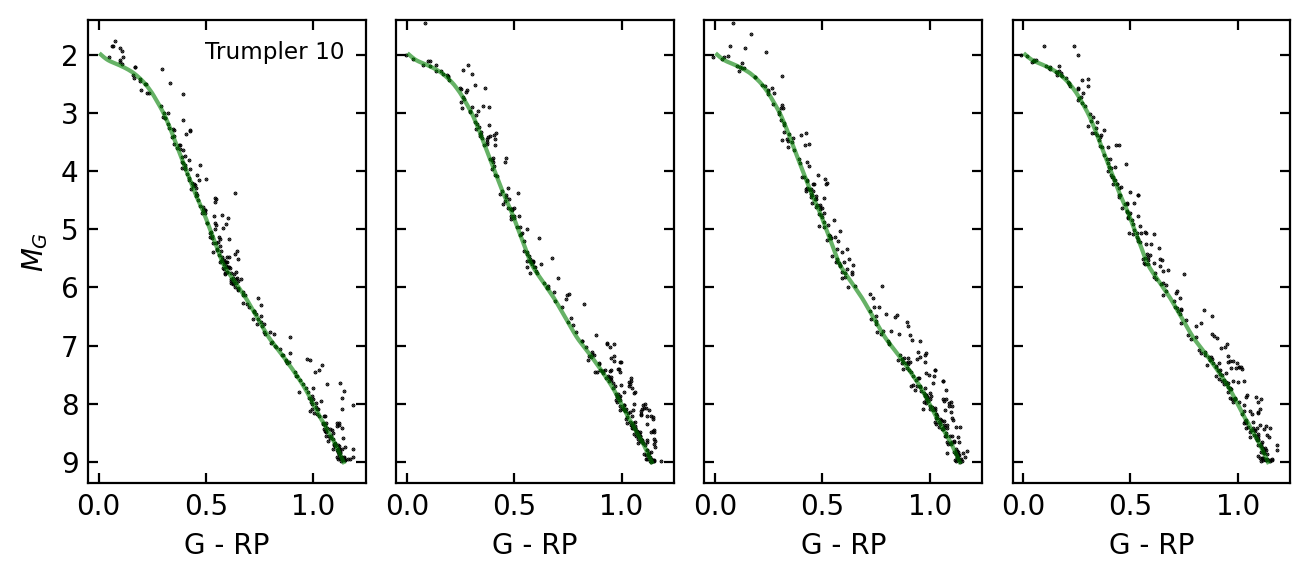}
    \end{subfigure}
    \hfill
    \begin{subfigure}[b]{0.95\textwidth}
        \centering
        \includegraphics[width=\columnwidth]{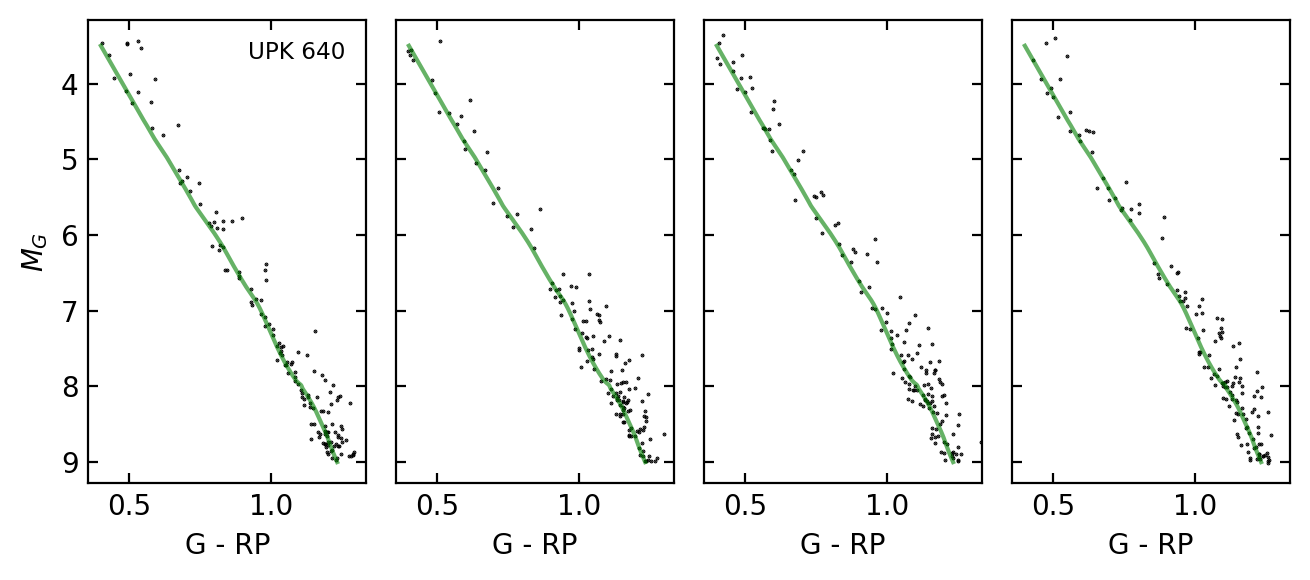}
    \end{subfigure}
    \caption{Observed CMD for each cluster, accompanied by 3 random realisations from the posterior maximum model for the Legendre 
    mass-ratio distribution representation.}
    \label{Realisation2}
\end{figure*}

In Figures \ref{Realisation} and \ref{Realisation2} the colour-magnitude diagrams from our data selection are presented alongside a randomly generated sample of three realisations for each cluster.


\section{Posterior Parameters}
In Table \ref{Posterior params} we include the values and 1 $\sigma$ uncertainties of each parameter. Each cluster was tested with different combinations of freeze parameters, and the blank spaces in the table indicate that the parameter was frozen for the best model of that cluster.

\begin{table*}
    \centering
    \caption[Posterior parameters of each cluster]{The posterior parameter values of each cluster along with 1 $\sigma$ uncertainties. Blank spaces indicate that the variable was frozen. \label{Posterior params}}
    \begin{tabular}{lcccccc}
         \hline
        & Collinder 69 & $\alpha$ Persei & Pleiades & NGC 6405 & Trumpler 10 & UPK 640 \\
         \hline\hline
         $\gamma$ 
         & $1.50^{+0.28}_{-0.27}$ & $1.10^{+0.25}_{-0.23}$ & 
         $2.65^{+3.02}_{-1.82}$ & $2.35^{+3.45}_{-2.22}$ &
         $5.16^{+0.80}_{-2.48}$ & $1.67^{+0.31}_{-0.29}$ \vspace{0.1cm}\\
         
         $c_0$ 
         &  &  &
         $5.06^{+3.97}_{-4.03}$ & $6.12^{+2.17}_{-4.04}$ &
         $0.49^{+1.45}_{-0.25}$ &  \vspace{0.1cm}\\
         
         $c_1$ 
         &  &  &
         $-3.28^{+2.60}_{-2.57}$ & $-4.57^{+2.91}_{-1.37}$ &
         $-0.27^{+0.14}_{-0.80}$ &  \vspace{0.1cm}\\

         $\sigma$ 
         & $1.29^{+0.60}_{-0.31}$ & $1.11^{+3.53}_{-0.11}$ & 
         $1.02^{+0.08}_{-0.02}$ & $1.03^{+0.10}_{-0.03}$ &
         $1.02^{+0.08}_{-0.02}$ & $1.16^{+3.55}_{-0.15}$ \vspace{0.1cm}\\
         
         $a_1$ 
         & $-0.68^{+0.35}_{-0.31}$ & $-0.84^{+1.51}_{-0.78}$ &
         $-1.67^{+0.25}_{-0.20}$ & $-1.48^{+0.24}_{-0.21}$ &
         $-1.61^{+0.25}_{-0.21}$ & $-1.02^{+1.23}_{-0.50}$ \vspace{0.1cm}\\
         
         $a_2$ 
         & $1.06^{+0.50}_{-0.48}$ & $1.00^{+1.08}_{-1.61}$ & 
         $2.43^{+0.26}_{-0.34}$ & $2.13^{+0.31}_{-0.39}$ & 
         $2.14^{+0.31}_{-0.34}$ & $1.17^{+0.90}_{-1.39}$ \vspace{0.1cm}\\
         
         $a_3$ 
         & $0.88^{+0.57}_{-0.57}$ & $-1.35^{+1.48}_{-0.89}$ & 
         $-2.07^{+0.39}_{-0.29}$ & $-2.14^{+0.36}_{-0.24}$ & 
         $-2.11^{+0.37}_{-0.27}$ & $-0.83^{+1.03}_{-1.12}$ \vspace{0.1cm}\\
         
         $a_4$ 
         & $0.14^{+0.67}_{-0.56}$ & $1.30^{+0.91}_{-1.34}$ & 
         $2.14^{+0.22}_{-0.37}$ & $1.98^{+0.29}_{-0.34}$ & 
         $2.19^{+0.22}_{-0.35}$ & $0.98^{+1,21}_{-1.03}$ \vspace{0.1cm}\\
         
         $a_5$ 
         & $0.13^{+0.74}_{-0.50}$ & $-0.09^{+0.56}_{-0.92}$ & 
         $-1.01^{+0.51}_{-0.44}$ & $-1.02^{+0.52}_{-0.45}$ & 
         $-1.06^{+0.51}_{-0.45}$ & $-0.13^{+0.46}_{-1.25}$ \vspace{0.1cm}\\
         
         $f_B$ 
         & $0.94^{+0.01}_{-0.05}$ & $0.65^{+0.20}_{-0.32}$ & 
         $0.93^{+0.02}_{-0.10}$ & $0.93^{+0.02}_{-0.10}$ & 
         $0.91^{+0.04}_{-0.12}$ & $0.77^{+0.17}_{-0.42}$ \vspace{0.1cm}\\
         
         $\dot{f_B}$ 
         &  & $-0.06^{+0.11}_{-0.05}$
         & $0.00^{+0.06}_{-0.03}$ &  $0.01^{+0.09}_{-0.03}$
         & $0.00^{+0.06}_{-0.05}$ & \vspace{0.1cm}\\
         
         $f_O$ 
         & $0.04^{+0.03}_{-0.02}$ & $0.04^{+0.04}_{-0.03}$ & 
         $0.02^{+0.02}_{-0.01}$ & $0.03^{+0.02}_{-0.02}$ & 
         $0.06^{+0.04}_{-0.03}$ & $0.02^{+0.03}_{-0.01}$ \vspace{0.1cm}\\
         
         $\log h$ 
         & $0.93^{+0.06}_{-0.08}$ & $0.76^{+0.10}_{-0.07}$ & 
         $0.73^{+0.05}_{-0.05}$ & $0.90^{+0.06}_{-0.06}$ & 
         $0.74^{+0.12}_{-0.09}$ & $0.97^{+0.02}_{-0.05}$ \vspace{0.1cm}\\

         $\log_{10}k$ 
         & $1.77^{+0.22}_{-0.25}$ &  &
          &  & 
          & $1.76^{+0.22}_{-0.25}$ \vspace{0.1cm}\\
         
         $M_0$ 
         & $0.13^{+0.05}_{-0.05}$ &  &
          &  &  & $0.10^{+0.05}_{-0.03}$
           \vspace{0.1cm}\\
         
         
         \hline
    \end{tabular}
\end{table*}

\section{Corner Plots}

In this section of the appendix we include in Figures \ref{fig:Col69 Corner}-\ref{fig:Upk640 Corner} the corner plots for each cluster using the best model. 

\begin{figure*}
    \centering
    \includegraphics[width=\textwidth]{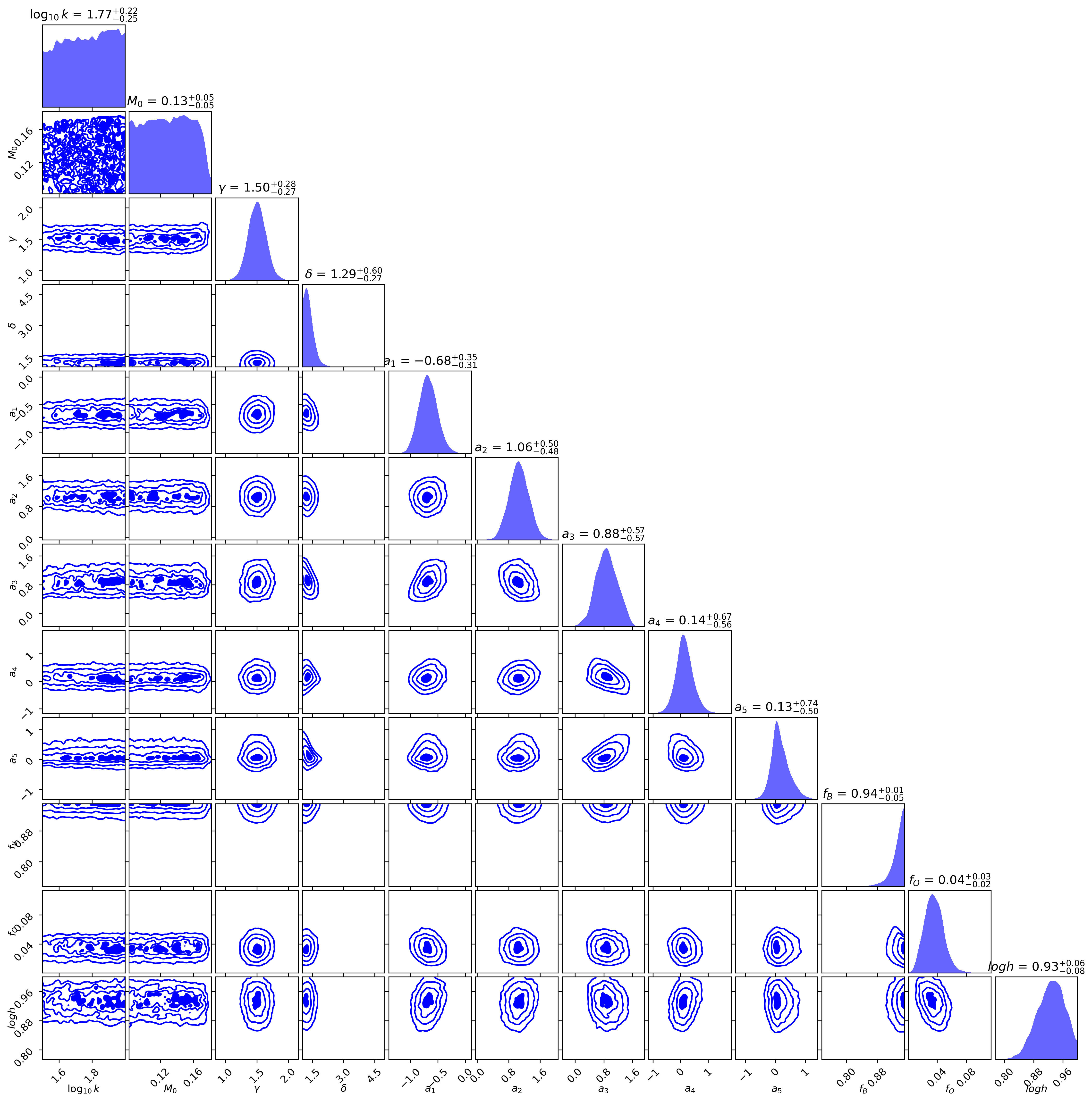}
    \caption{Corner plot of Collinder 69.}
    \label{fig:Col69 Corner}
\end{figure*}

\begin{figure*}
    \centering
    \includegraphics[width=\textwidth]{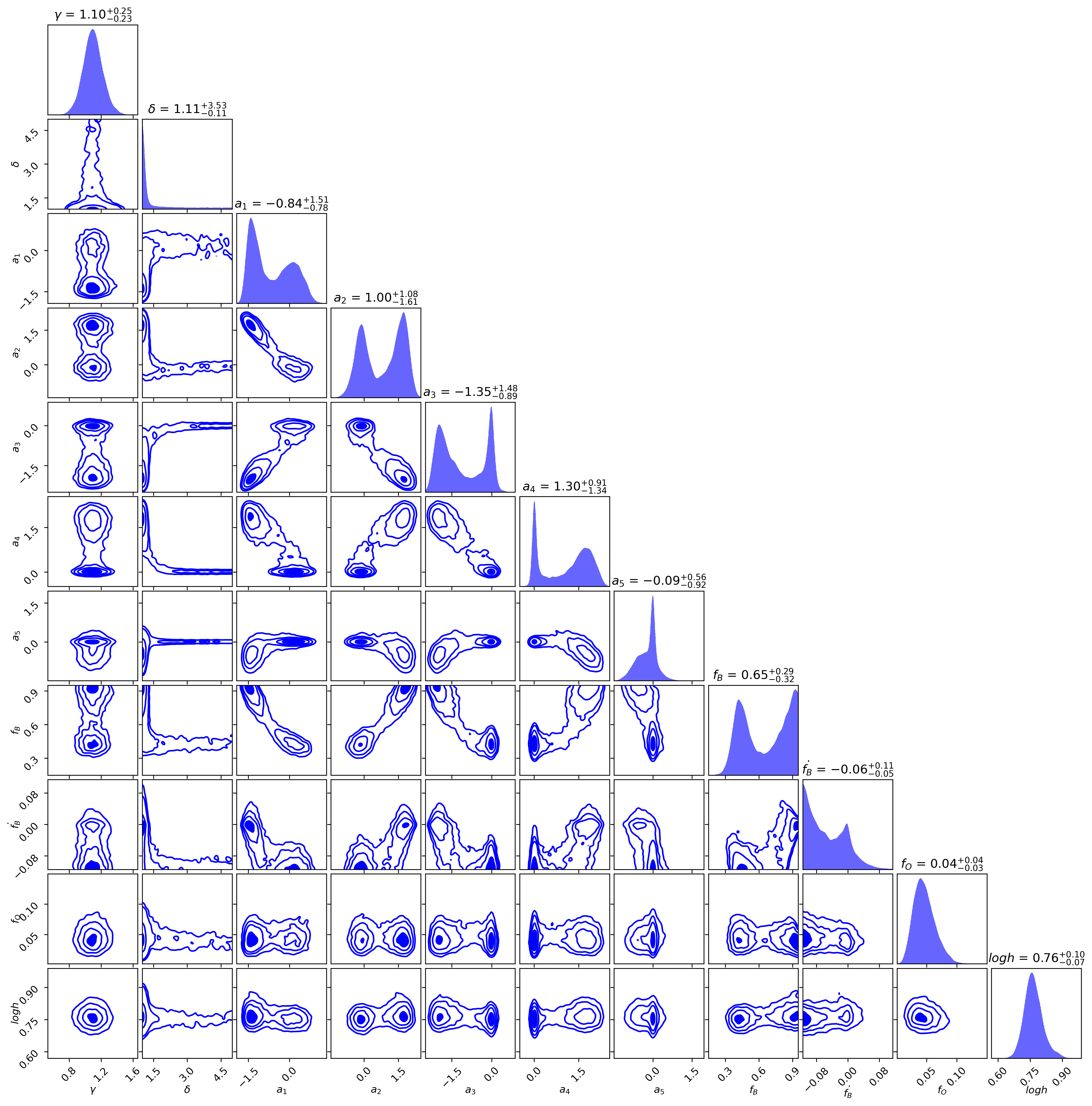}
    \caption{Corner plot of $\alpha$ Persei.}
    \label{fig:Mel20 Corner}
\end{figure*}

\begin{figure*}
    \centering
    \includegraphics[width=\textwidth]{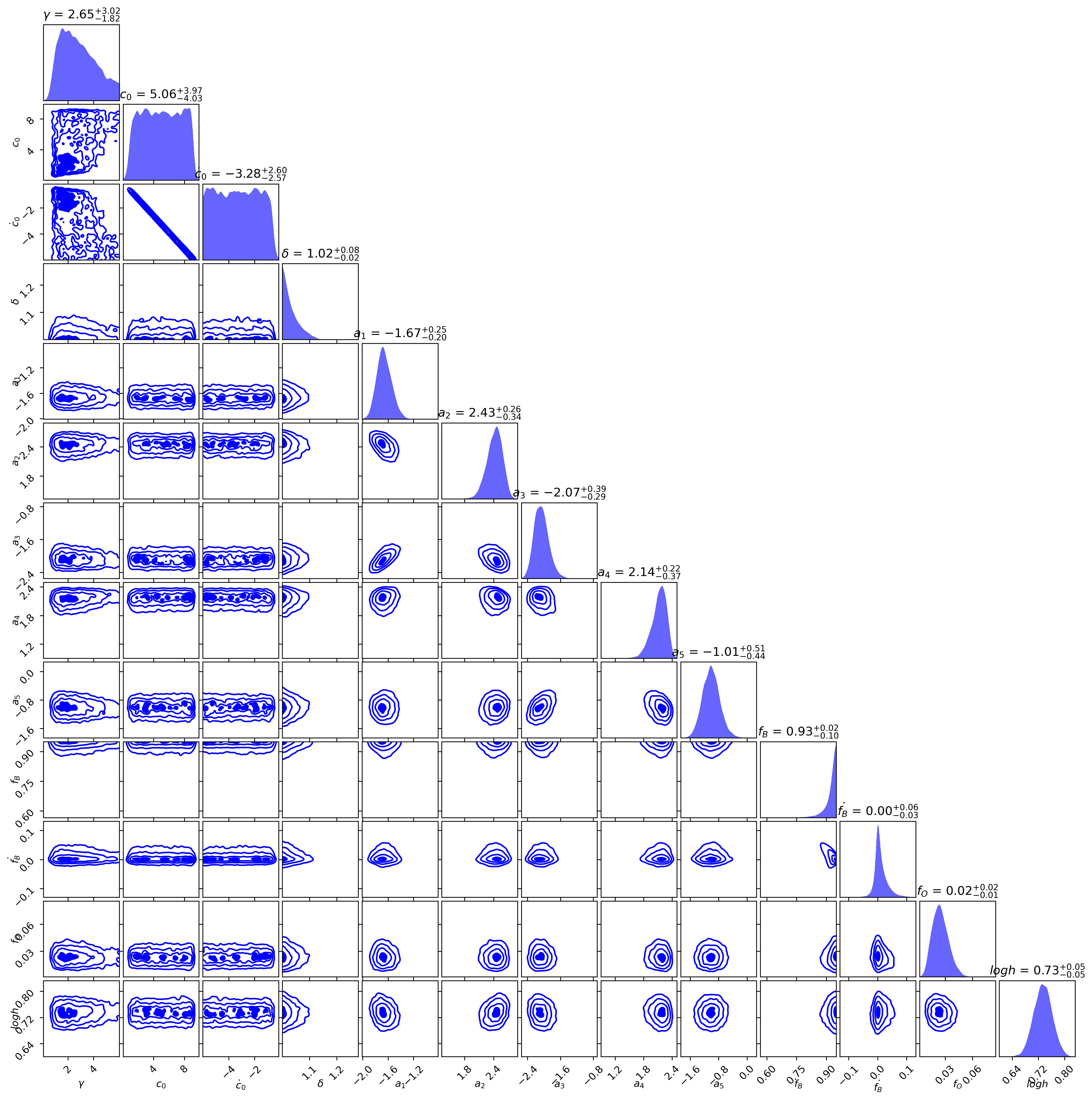}
    \caption{Corner plot of Pleiades.}
    \label{fig:Mel22 Corner}
\end{figure*}

\begin{figure*}
    \centering
    \includegraphics[width=\textwidth]{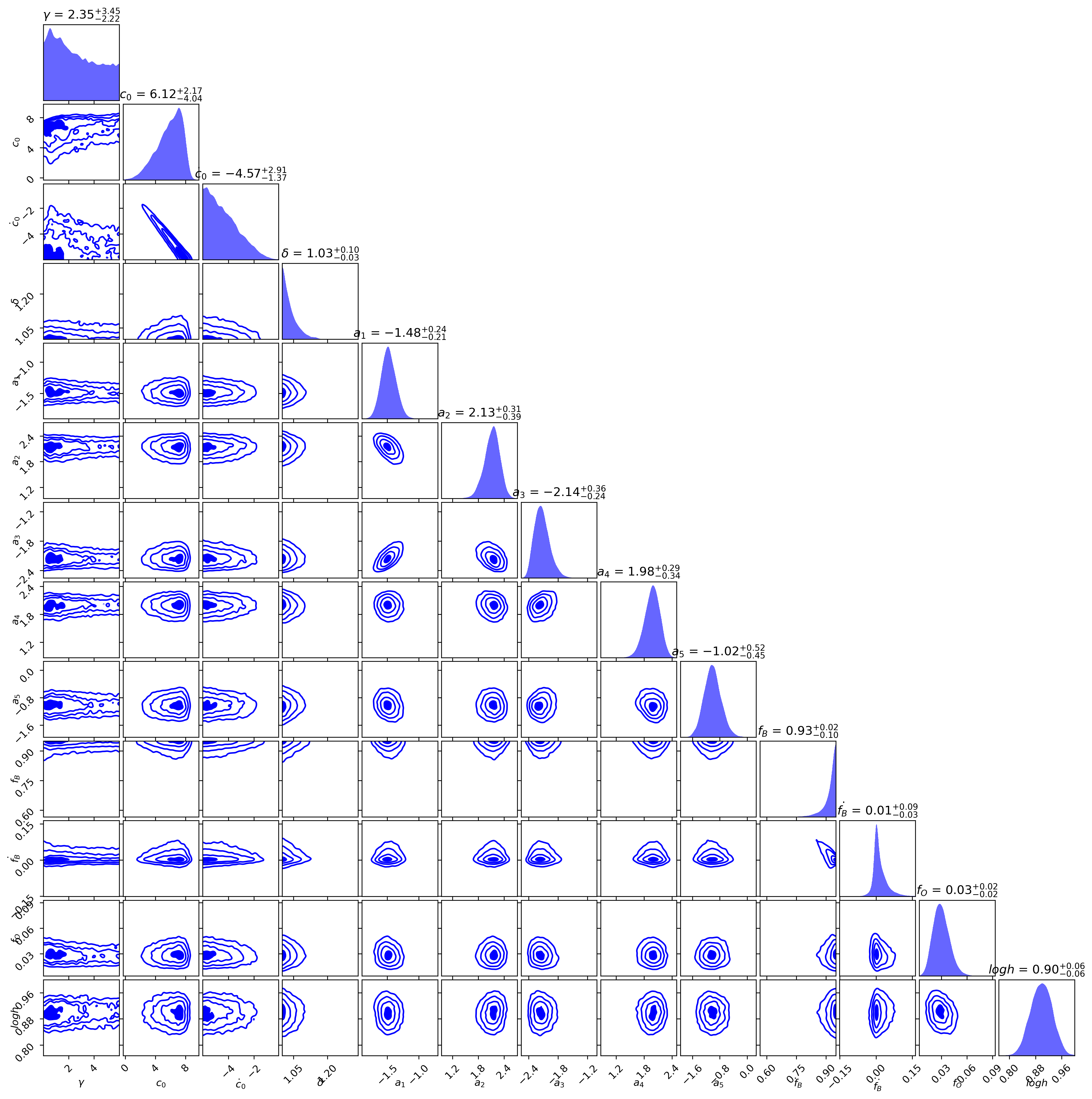}
    \caption{Corner plot of NGC 6405.}
    \label{fig:NGC6405 Corner}
\end{figure*}

\begin{figure*}
    \centering
    \includegraphics[width=\textwidth]{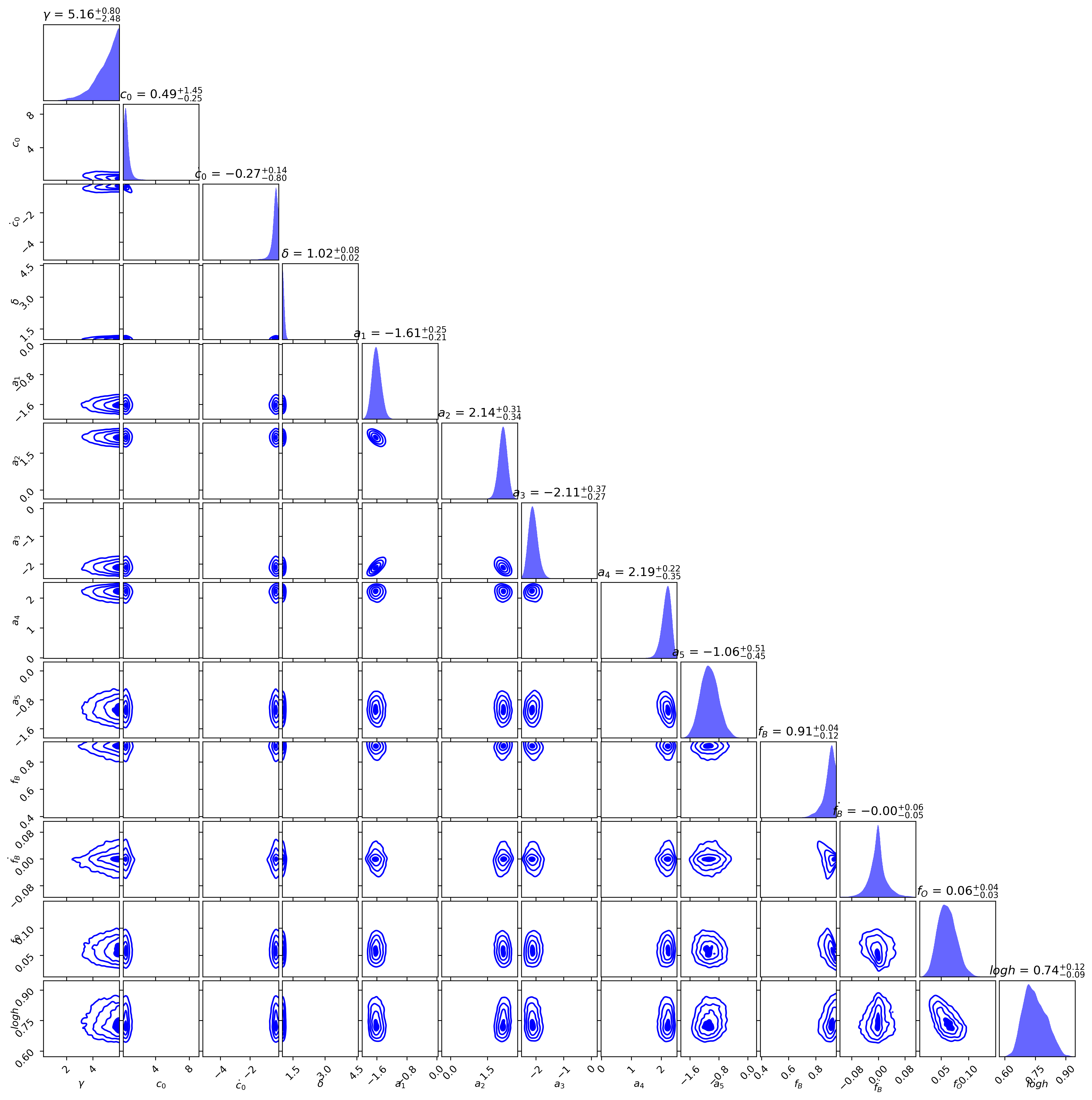}
    \caption{Corner plot of Trumpler 10.}
    \label{fig:Tr10 Corner}
\end{figure*}

\begin{figure*}
    \centering
    \includegraphics[width=\textwidth]{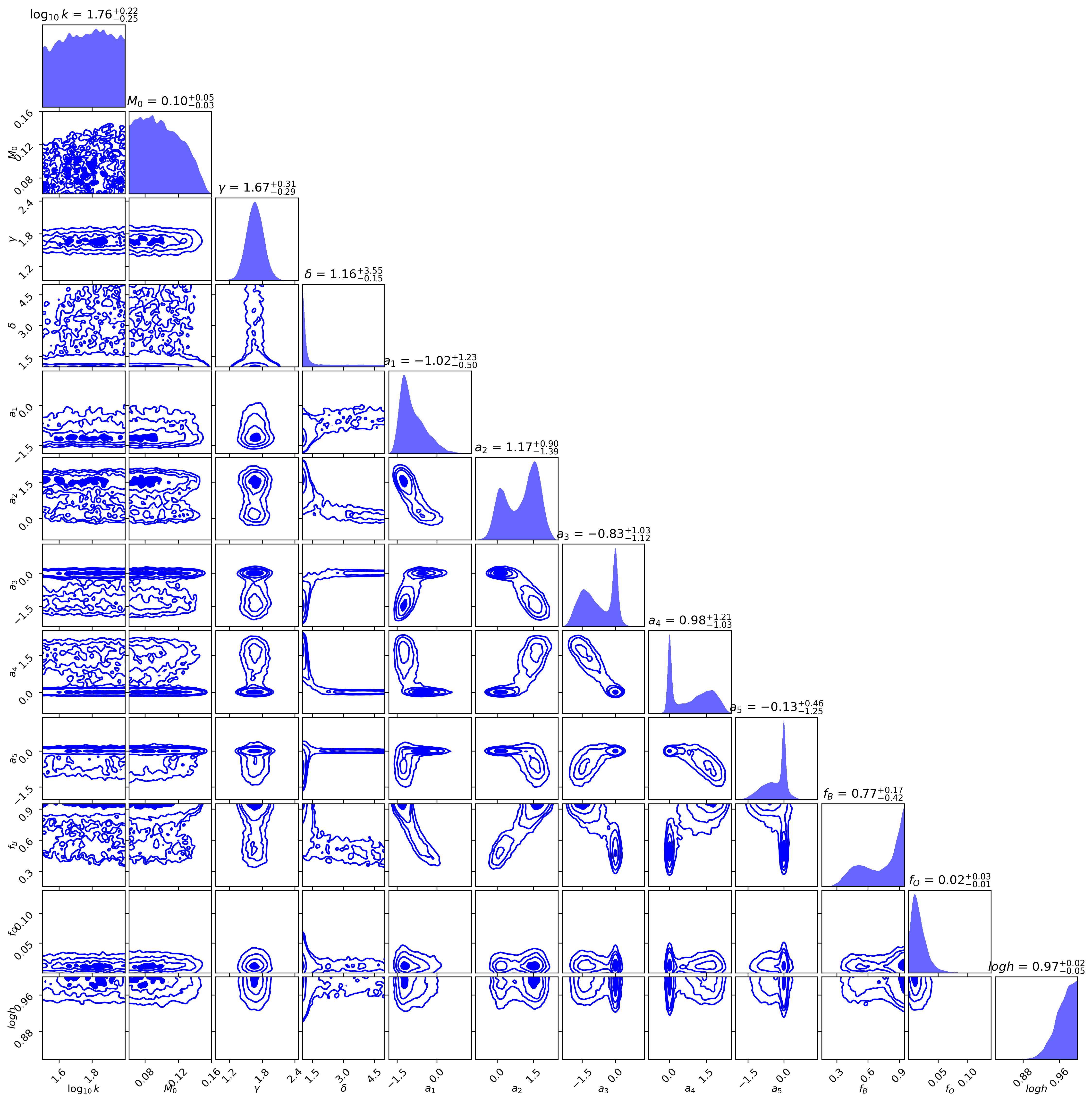}
    \caption{Corner plot of UPK 640.}
    \label{fig:Upk640 Corner}
\end{figure*}

\section{Coefficients of the linear fits to parameter relationships}\label{Coeffcient Relationships}

In Table \ref{Coeffcient Relationships} we include the coefficients of the linear lines that best fit the data from Figure \ref{LH4_10Bins}. The lines are defined as \begin{equation}
    B_0x + B_1.
\end{equation}

\begin{table}
    \caption{Coefficients of the equation $B_0x+B_1$ displayed as $[B_0,B_1]$ for the linear fits in Figure \ref{LH4_10Bins}.}
    \label{Coefficients Table Lines}
    \begin{tabular}{cccc}
         \hline
         & \multicolumn{3}{c}{Mass-ratio Distribution Model} \\
         Parameters & Legendre & Histogram & Histogram\\
         &&(4 Bins)&(10 Bins)\\
         \hline\hline
         $f_B(q\geq0.5)$  & \multirow{2}{*}{[3.921, -0.271]} & \multirow{2}{*}{[9.395, -1.185]} & \multirow{2}{*}{[5.275, -0.446]} \\
         and $\mathrm{FQ_{75}}$ &&&\vspace{0.05cm}\\
         Age and  & \multirow{2}{*}{[-0.069, 0.736]} & \multirow{2}{*}{[-0.056, 0.610]} & \multirow{2}{*}{[-0.095, 0.910]} \\
         $f_B(q\geq0.5)$ &&&\vspace{0.05cm}\\
         \multirow{2}{*}{Age and $\mathrm{FQ_{75}}$} & \multirow{2}{*}{[-0.263, 2.556]} & \multirow{2}{*}{[-0.494, 4.325]} & \multirow{2}{*}{[-0.335, 3.086]} \\\\
         \hline
    \end{tabular}
\end{table}

\bsp	
\label{lastpage}
\end{document}